\documentclass[10pt]{iopart}

\usepackage{iopams}
\usepackage{amsgen, amsfonts, amsbsy, amssymb}
  \expandafter\let\csname equation*\endcsname\relax
  \expandafter\let\csname endequation*\endcsname\relax
\usepackage{amsmath,amssymb,bm} 
\usepackage{csquotes}   
\usepackage{siunitx}
\usepackage{mathrsfs}
\usepackage{graphicx}
\usepackage{url}
\usepackage{dsfont} 
\usepackage{hyperref}
\usepackage{cite}   

\newcommand{\beqref}[1]{{(\ref{#1})}}
\renewcommand{\eqref}[1]{{equation~\beqref{#1}}}

\newcommand{\figref}[1]{{figure~\ref{#1}}}
\newcommand{\Figref}[1]{{Figure~\ref{#1}}}

\newcommand{\E}{{\bf E}}        
\newcommand{\D}{{\bf D}}        
\newcommand{\Bfield}{{\bf B}}        
\newcommand{\Hfield}{{\bf H}}        
\newcommand{\NVF}{{\bf N}}   
\newcommand{\errsym}{{\textit{e}}}
\newcommand{\Errsym}[1]{{\textit{e}\big\{ #1 \big\}}}
\newcommand{\x}{{\bf x}}    
\newcommand{\n}{{\bf n}}    

\usepackage{stmaryrd} 
\usepackage{nicefrac}
\usepackage{minibox}
\newcommand{\fmatrix}[1]{{\llbracket #1 \rrbracket}}
\newcommand{\fvec}[1]{{[#1]}}
\newcommand{\rec}[1]{{ {\mathcal R}\left(#1\right)}}
\newcommand{\idmat}{{ \mathds 1}}
\newcommand{\grnom}[1]{{ \left\|#1\right\|_G}}  

\def\josa{J.\ Opt.\ Soc.\ Am.\ }
\def\josaa{J.\ Opt.\ Soc.\ Am.\ A.\ }
\def\josab{J.\ Opt.\ Soc.\ Am.\ B.\ }

\def\nm{Nat.\ Mater.\ }
\def\nl{Nano\ Lett.\ }
\def\nt{Nanotechnol.\ }
\def\sab{Sens.\ Act.\ B:\ Chem.\ }
\def\oe{Opt.\ Express\ }

\def\prb{Phys.\ Rev.\ B\ }
\def\prl{Phys.\ Rev.\ Lett.\ }

\begin{document}

\title[Accurate near-field calculation in the rigorous coupled-wave analysis method]{Accurate near-field calculation in the rigorous coupled-wave analysis method}

\author{Martin Weismann$^{1,2}$, Dominic F G Gallagher$^{2}$ and Nicolae~C~Panoiu$^{1}$}

\address{$^1$ Department of Electronic and Electrical Engineering, University College London, Torrington Place, WC1E 7JE, London, UK}
\address{$^2$ Photon Design Ltd, 34 Leopold Street, OX4 1TW, Oxford, UK}
\ead{martin.weismann.12@ucl.ac.uk}
\vspace{10pt}
\begin{indented}
\item[]{\today}
\end{indented}

\begin{abstract}
The rigorous coupled-wave analysis (RCWA) is one of the most successful and widely used methods
for modeling periodic optical structures. It yields fast convergence of the electromagnetic
far-field and has been adapted to model various optical devices and wave configurations. In this
article, we investigate the accuracy with which the electromagnetic near-field can be calculated
by using RCWA and explain the observed slow convergence and numerical artifacts from which it
suffers, namely unphysical oscillations at material boundaries due to the Gibb's phenomenon. In
order to alleviate these shortcomings, we also introduce a mathematical formulation for accurate
near-field calculation in RCWA, for one- and two-dimensional straight and slanted diffraction
gratings. This accurate near-field computational approach is tested and evaluated for several
representative test-structures and configurations in order to illustrate the advantages provided
by the proposed modified formulation of the RCWA.
\end{abstract}

%
\vspace{2pc}

\submitto{\JOPT}
%
%
\ioptwocol

\section{Introduction}
\label{sec:introduction} The study of the interaction of electromagnetic waves with matter has
spawned a large variety of methods to analytically or numerically solve the Maxwell equations
(MEs). The results obtained from those methods are invaluable for understanding, validating,
predicting, and guiding experimental efforts and for the design process of electromagnetic and
optical devices. Amongst the numerical methods used in computational electromagnetics, one can
fundamentally distinguish between time-domain methods, which directly incorporate the transient
behavior of electromagnetic waves, such as the finite-difference time-domain method
\cite{taflove2000computational}, and frequency-domain methods, like the finite-element
frequency-domain method \cite{jin14wiley}, which directly determine time-harmonic solutions to
MEs. The former methods are general purpose methods and are capable of simulating virtually any
electromagnetic structure comprising metallic or dielectric objects of arbitrary size and shape.
However, for certain structures and applications, other methods can be superior in terms of
runtime and accuracy. For example, the multiple scattering method \cite{martin06cup} is a series
expansion method in the frequency domain that is tailored to calculate interaction of light and
clusters of spherical particles \cite{mtm96jqsrt} or cylindrical rods \cite{ftm94josaa}, and as
such it is superior to other methods when applied to these particular geometries.

An important field of applications, where several specialized numerical algorithms exist
\cite{popov12popov}, is the study of periodic optical structures, including diffraction gratings
\cite{palmer05newport} and periodic metamaterials \cite{cui09metamaterials}. Amongst the numerical
methods for periodic structures, such as the differential theory of gratings \cite{np02tf}, the
$\mathcal{C}$-method \cite{chandezon1982multicoated}, integral methods \cite{popov12popov,
Maystre78JOSA}, and the generalized source method \cite{st12jqsrt,wgp15josab}, the most widely
used method is the rigorous coupled-wave analysis (RCWA)
\cite{Moharam81josaa,Moharam86josaa,mgp95josaa}. It is a modal method in the frequency domain and
is based on the decomposition of the periodic structure and the pseudo-periodic solution of MEs in
terms of their Fourier series (FS) expansion, hence the periodicity is naturally incorporated into
the numerical method. RCWA was initially developed for modeling one-dimensional (1D) diffraction
gratings, but with the introduction of fast converging Fourier factorization rules for modeling 1D
\cite{lm96josaa,Li1996josaa} and two-dimensional (2D) \cite{l97josaa,eb10oe,srk07josaa}
structures, its formulation for isotropic and anisotropic materials \cite{li03josaa} as well as
multilayered \cite{Moharam95josaa2} and oblique structures \cite{gt10josaa}, RCWA has evolved to
describe arbitrary, 2D-periodic structures. The method has been successfully applied to model
diffraction gratings, diffractive optical elements, surface coatings, spectroscopic applications,
photonic crystals, and periodic metamaterials. It should be noted, that in most cases the
functionality of these applications of periodic structures depends on the electromagnetic
far-field, i.e. the propagating diffraction orders, and it is known that the RCWA delivers fast
converging and accurate far-field results, as reported in many works
\cite{Moharam81josaa,Moharam86josaa,mgp95josaa,lm96josaa,Li1996josaa,l97josaa,eb10oe,srk07josaa,li03josaa,Moharam95josaa2,gt10josaa}.

There is, however, a range of novel applications, which rely on the optical near-field of a
periodic structure, especially at their surface, such as surface-enhanced Raman spectroscopy
\cite{kwk97prl}, surface second-harmonic generation \cite{nsh06prl,cpo07prb,bp10prb}, and
near-field sensing \cite{hyg99sab,kep09natm,lmw10nl,bp11nt}. These applications require a suitably
designed near-field distribution, usually optimized for maximum field enhancement within specific
spatial domains. Although there have been significant advances in experimental optical near-field
measurement techniques \cite{adam08oe}, these techniques are still in their development state and
not readily available to accurately characterize complex photonic nanostructures. These
applications and experimental shortcomings lead to a critical demand for numerical methods for
periodic structures that can facilitate an accurate calculation of electromagnetic near-fields,
and more importantly, the design of gratings with optimized near-field patterns. With very few
exceptions \cite{popov2002staircase, eb10oe}, a thorough investigation of numerically calculated
near-fields in the RCWA has been largely neglected during the development of the method, as it is
often merely considered a post-processing step. Moreover, additional reasons for the scarcity of
reports on the convergence and the accuracy of the numerically computed electromagnetic near-field
in RCWA, are the slow near-field convergence and spurious oscillations displayed by these fields
\cite{popov2002staircase}.

In this paper we address the issue of inaccurate near-field calculations in the RCWA in several
ways. First, we use some generic cases of diffraction gratings to illustrate the slow convergence
of RCWA for near-field calculations and reveal the reasons for this behavior. Based on this
analysis and the continuity properties of the electromagnetic fields, a new formulation of the
field evaluation is proposed, which yields faster convergence of the electromagnetic near-fields
and explicitly fulfills the continuity properties of the electromagnetic field at material
interfaces. This improved formulation of the numerical evaluation of the near-fields is then
benchmarked against the current formulation, with the aim of making RCWA an effective numerical
method for modeling modern nanophotonic applications that rely on highly accurate near-field
calculations.

The remainder of the paper is organized as follows: Section~\ref{sec:1D} will introduce the reader
to the problem of inaccurate/spurious near-fields in RCWA for 1D structures and explain the
overall strategy for the accurate field evaluation. Section~\ref{sec:Formulation} will extend the
ideas gained from the analysis of 1D structures to arbitrary, straight or obliquely etched 2D
gratings and will present the underlying mathematical formalism. Then in section~\ref{sec:2D},
computational results for 2D gratings will be presented and discussed. Section~\ref{sec:1D2Dz}
will investigate near- and far-field convergence of the modified method for slanted gratings,
before final conclusions about the capability of the improved RCWA for modeling electromagnetic
near-fields will be drawn in section~\ref{sec:Conclusions}.

\section{Accurate near-field evaluation for 1D-periodic structures}
\label{sec:1D} Although the last major roadblock that was precluding RCWA from becoming a highly
effective method for modeling 1D periodic structures has been removed when the correct Fourier
factorization rules for transverse magnetic (TM) polarization were introduced
\cite{lm96josaa,Li1996josaa}, a few topics have continued to attract attention, namely the
convergence for slanted 1D-periodic gratings \cite{gt10josaa} and the numerical instabilities
associated to highly-conductive gratings \cite{lpt07josaa}. These improvements and refinements of
the method are concerned with the accuracy of the far-field calculation and they achieve excellent
convergence results for the coefficients of and power carried by the diffraction orders in the
cover and substrate regions of a grating, namely the far-field. This picture changes if the
near-field is considered. Thus, in this section we investigate the numerically calculated
near-field inside the periodic grating region via RCWA, and explain its slow convergence and the
origin of the observed high spatial frequency oscillations. We also propose an improved field
evaluation approach, which yields faster converging near-field profiles free of numerical
artifacts.

\begin{figure}[t]
\centering \includegraphics[width=\linewidth]{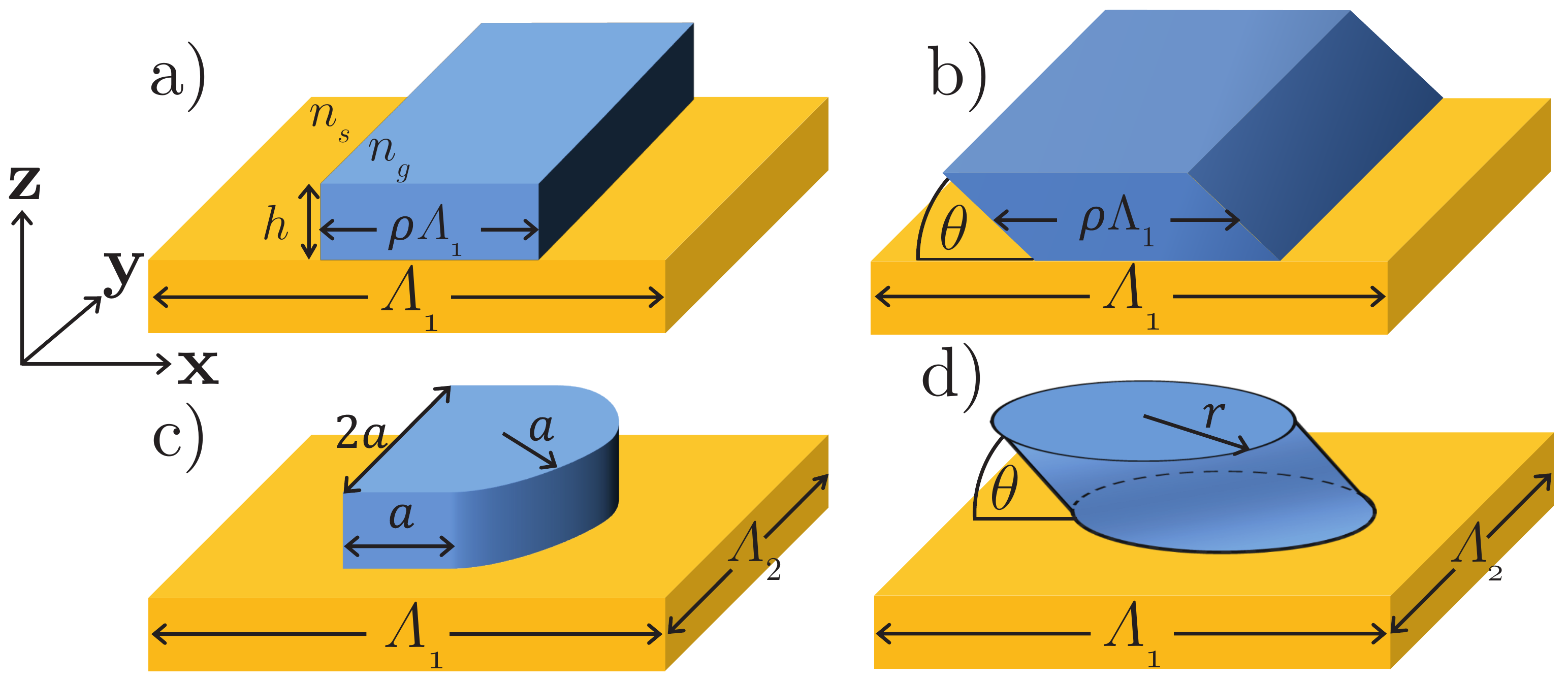} \caption{Grating structures mounted on a
homogeneous substrate (refractive index, $n_s$) considered in this study: a) 1D binary grating,
filling factor, $\rho$; b) 1D binary grating, slanted by $\theta=\pi/4$ w.r.t. the $x$-axis. c) 2D
grating with rectangle-semi-circular cross-section. d) 2D cylindrical grating, slanted by
$\theta=\pi/4$ w.r.t. the $x$-axis. The height of all gratings is denoted with $h$, their period
is $\Lambda_1$ (and $\Lambda_2$ for 2D-periodic gratings), and their refractive index is denoted
by $n_g$. The cover medium is vacuum with $n_c=1$.} \label{fig:Structures}
\end{figure}

The ideas developed in this paper are applicable to any periodic grating structure with sharp
boundaries, but in order to conduct a comprehensive assessment of the accuracy of the near-field
calculations achievable with the improved RCWA introduced in this work, a number of generic test
structures (1D- and 2D-periodic, straight and slanted) have been chosen, as per
\figref{fig:Structures}. To widen the spectrum of test configurations, three different materials
are considered for the grating: silica ($\mathrm{SiO_{2}}$) as a dielectric with low index of
refraction ($n_g=n_{SiO_{2}}=1.45$), silicon (Si) as a high refractive index dielectric
($n_g=n_{Si}=3.4$) and a metal, gold (Au), with index of refraction $n_g=n_{Au}=0.97+1.87i$
\cite{jc72prb}. In all examples the substrate is $\mathrm{SiO_{2}}$ ($n_s=n_{SiO_{2}}$). Normal
incidence and TM-polarization is considered in all of the sections, i.e. the incident electric
field amplitude is oriented along the $x$-axis: $\E_{inc}(\x)=(1,0,0)^TE_{inc}(\x)$.

The example considered in this section is depicted in \figref{fig:Structures}(a). It consists of a
binary grating with period, $\Lambda_1=\SI{1}{\micro\meter}$, height, $h=\SI{0.25}{\micro\meter}$,
and filling factor, $\rho=0.5$. The grating is illuminated by a normally incident plane wave with
wavelength, $\lambda=\SI{0.51}{\micro\meter}$.

As the first step of our investigation, we used RCWA to calculate the far-field. In order to
quantify the convergence of the RCWA method, the relative error of the far-field is defined as:
\begin{align}
\errsym_F(N) = \frac{\sqrt{|T^{(N)}-T^{ref}|^2 + |R^{(N)}-R^{ref}|^2}}{\sqrt{|T^{ref}|^2 +
|R^{ref}|^2}}, \label{eq:defFarFieldError}
\end{align}
where $T^{(N)}$ ($R^{(N)}$) denotes the relative transmitted (reflected) power corresponding to a
discretization with $2N+1$ complex FS coefficients. The discretization parameter, $N$, represents
the number of harmonics retained for each dimension. Moreover, $T^{ref}$($R^{ref}$) is a reference
value that is considered to be the exact solution or a sufficiently good approximation of the
exact solution. Due to the absence of the exact solution of the diffraction grating problem, the
reference values are chosen to be numerical values obtained by high-$N$ simulations; in our case
we chose $T^{ref}=T^{(905)}$ and $R^{ref}=R^{(905)}$.

\begin{figure}[t]
\centering \includegraphics[width=\linewidth]{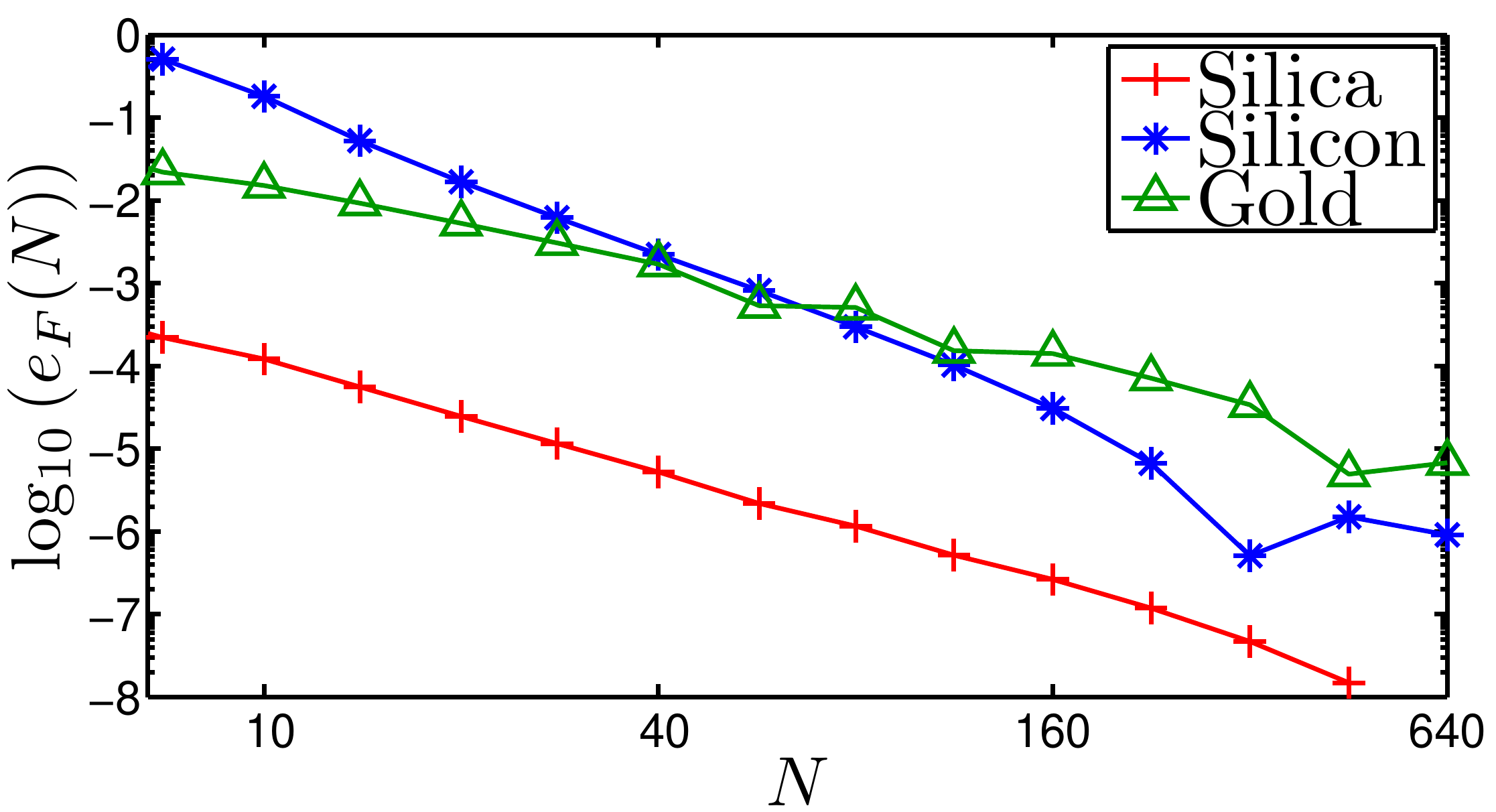} \caption{Far-field relative error vs. the
number of harmonics, determined for three different materials.} \label{fig:SelfConvergence1D}
\end{figure}

The far-field relative error, $\errsym_F(N)$, for increasing number of harmonics,
$N=5,\ldots,640$, is depicted in \figref{fig:SelfConvergence1D}. As this figure illustrates,
RCWA converges quickly for all three materials. Specifically, in order to achieve a self-error of
$\errsym_F(N)< 1\%$ (as a generic criterion adopted here for an accurate far-field calculation),
for the three materials, silica (red crosses), silicon (blue stars), and gold (green triangles), a
relatively small number of harmonics is necessary, namely $N>5$, $N>25$, and $N>13$, respectively.

As mentioned in the introductory section, the far-field of optical gratings and periodic
structures has been the physical quantity of most interest from experimental point of view, hence
the characterization of a numerical method by means of the far-field convergence has usually been
the adopted strategy. This approach, however, largely neglects the electromagnetic near-field
predicted by a specific method. On the other hand, the near-field is of fundamental importance for
modeling plasmonic effects or optical nonlinear phenomena in devices with size comparable to or
smaller than the operating wavelength, effects whose description relies on accurate calculations
of the electromagnetic near-field.
\begin{figure}[htbp]
\centering \includegraphics[width=\linewidth]{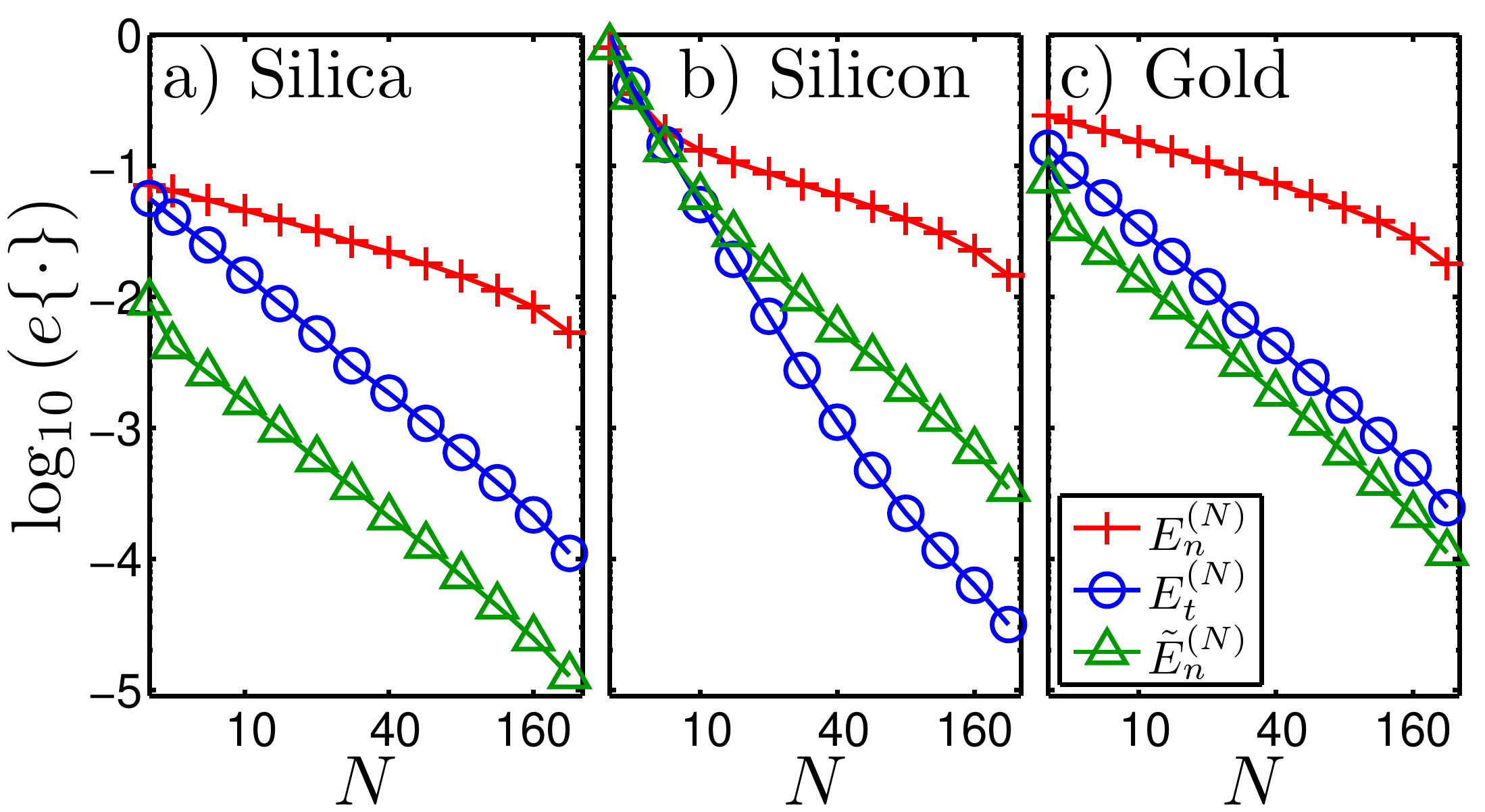} \caption{Self convergence of the tangential
and normal electric field components $E_z$ and $E_x$ and $\tilde E_x$ inside the grating structure
described by their self-errors $\Delta E_t(N)$ (green triangles), $\Delta E_n(N)$ (red crosses)
and $\Delta \tilde E_n(N)$ (blue circles), respectively for the three benchmark structures made of
Silica (a), Silicon (b) and Gold (c). } \label{fig:SelfConvergenceField1D}
\end{figure}

In order to characterize the numerically obtained near-fields, we define the grating norm,
$\grnom{\cdot}$, of a scalar or vectorial function, $f$, in the grating region as follows:
\begin{align}
\grnom{f}  &= \left(\int_{0}^{h}\int_{-\Lambda_1/2}^{\Lambda_1/2} \vert f(x,z)\vert^2\ d x \ d z
\right)^{\nicefrac 12}, \label{eq:NormG}
\end{align}
where the $z$-integration extends over the bulk of the periodic region. The grating norm is used
to define the near-field error, $\Errsym{E_\alpha^{(N)}}$, of the scalar field components
$E^{(N)}_\alpha$, $\alpha=x,y,z$, of a near-field, $\E^{(N)}$, numerically obtained using $N$
harmonics:
\begin{align}
    \Errsym{ E_\alpha^{(N)}} &= \grnom{E_\alpha^{(N)} - E_\alpha^{ref}} \Big/  \grnom{E_\alpha^{ref}}.  \label{eq:DeltaEalpha}
\end{align}
Here, $E_\alpha^{(N)}$ denotes the FS reconstruction given by the $2N+1$ central Fourier
coefficients, $E_{\alpha n}$, which are calculated by RCWA:
\begin{align*}
  E_\alpha^{(N)}(x) &= \sum_{n=-N}^N E_{\alpha n} \exp\left(i n \frac{2\pi}{\Lambda_1}x \right).
\end{align*}

Similarly to the far-field calculations, $\E^{ref}=\E^{(905)}$ is obtained by a high-resolution
RCWA simulation with $N=905$. The near-field self-convergence for the tangential component, $E_t =
E_z$, and the normal component, $E_n = E_x$, of the electric field is depicted in
\figref{fig:SelfConvergenceField1D}. For all three materials, the self-convergence of $E_t$
(blue circles) is fast and comparable to the far-field self-convergence (see
\figref{fig:SelfConvergence1D}). The normal component, $E_n$ (red crosses), however, exhibits
much slower convergence and even at the highest numerical resolution of $N=640$ a relative error
of $\Errsym{E_x^{(640)}}>0.9\%$ still remains, whereas the error of the tangential field,
$\Errsym{E_z^{(640)}}<0.08\%$, for all materials.

One intriguing question raised by the data plotted in figures~\ref{fig:SelfConvergence1D} and
\ref{fig:SelfConvergenceField1D} is why the normal component of the near-field converges much more
slowly than the far-field power and the tangential component of the near-field. Or put it the
other way around: how can the far-field converge quickly when the near-fields have not converged
yet? There are two factors that explain this behavior: {\itshape i)} The far-field consists of a
superposition of a small number of propagating diffraction orders, i.e., plane waves. Hence, the
far-field requires a small number of FS components to be reconstructed and does not suffer from
Gibb's phenomenon. {\itshape ii)} RCWA does not depend on the representation of the discontinuous
normal field $E_n$. Instead, the method relies on the correct Fourier factorization of the
continuous normal component of the displacement field, $D_n$, which can be accurately described by
a FS, i.e. without spurious oscillations. The second observation is at the core of the accurate
near-field calculation that is introduced in this work. Specifically, instead of reconstructing a
discontinuous physical quantity, i.e. $E_n(\x)$, directly from its FS coefficients, it is more
effective to reconstruct a continuous quantity, the normal component of the displacement field,
$D_n(\x)$, and then divide it by a discontinuous quantity, the electric permittivity,
$\varepsilon_0\varepsilon(\x)$, which is known in the space-domain where one seeks to solve the
diffraction problem.

For 1D-periodic gratings, we define the modified normal component of the electric field:
\begin{align}
    \tilde E^{(N)}_n(\x) = \varepsilon_0^{-1}D^{(N)}_n(\x)/\varepsilon(\x), \label{eq:ModEn1D}
\end{align}
where $\varepsilon_0$ is the vacuum permittivity and $\varepsilon(\x)$ the relative permittivity
at position, $\x$. Note that $E^{(N)}_n(\x)$ and $\tilde E^{(N)}_n(\x)$ represent the same
physical quantity, namely the normal component of the electric field. However, whereas the former
is found by using RCWA to solve directly for the electric field, the latter one is determined by
first calculating the displacement field and then the electric field via \eqref{eq:ModEn1D}.

The error, $\Errsym{\tilde E_n^{(N)}}$, of the modified normal component, $\tilde E_n$, is shown
in \figref{fig:SelfConvergenceField1D} and encoded in green triangle. It is found to converge as
fast as the fast-convergent tangential component, $E_t$, and as fast as the power in the far-field
shown in \figref{fig:SelfConvergence1D}. Even at the highest considered resolution, $N=640$, the
conventional formulation of the RCWA yields a self error of $\Errsym{E_n^{(640)}}>9\cdot 10^{-3}$
for all three materials. By contrast, the same self error $\Errsym{\tilde
E_n^{(N)}}<9\cdot10^{-3}$ of the modified normal field $\tilde E_n$ can be achieved by using as
few as $N=70$ harmonics.


\begin{figure}[bt]
\centering
\minibox{
\includegraphics[width=\linewidth]{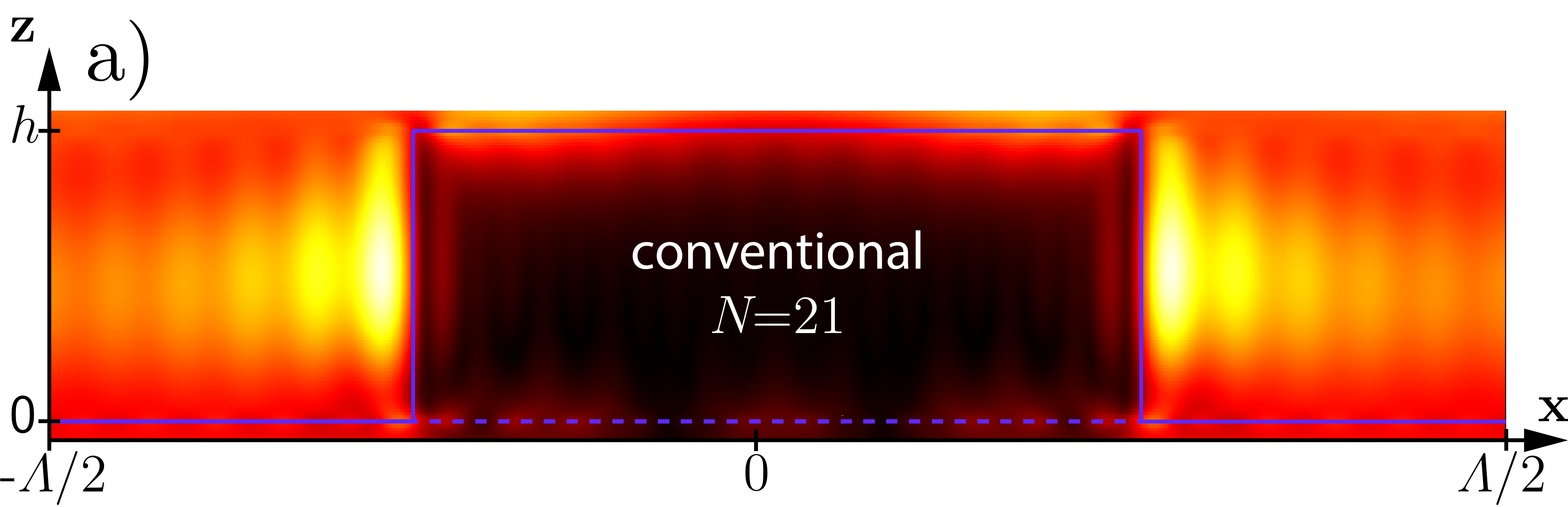}\\
\includegraphics[width=\linewidth]{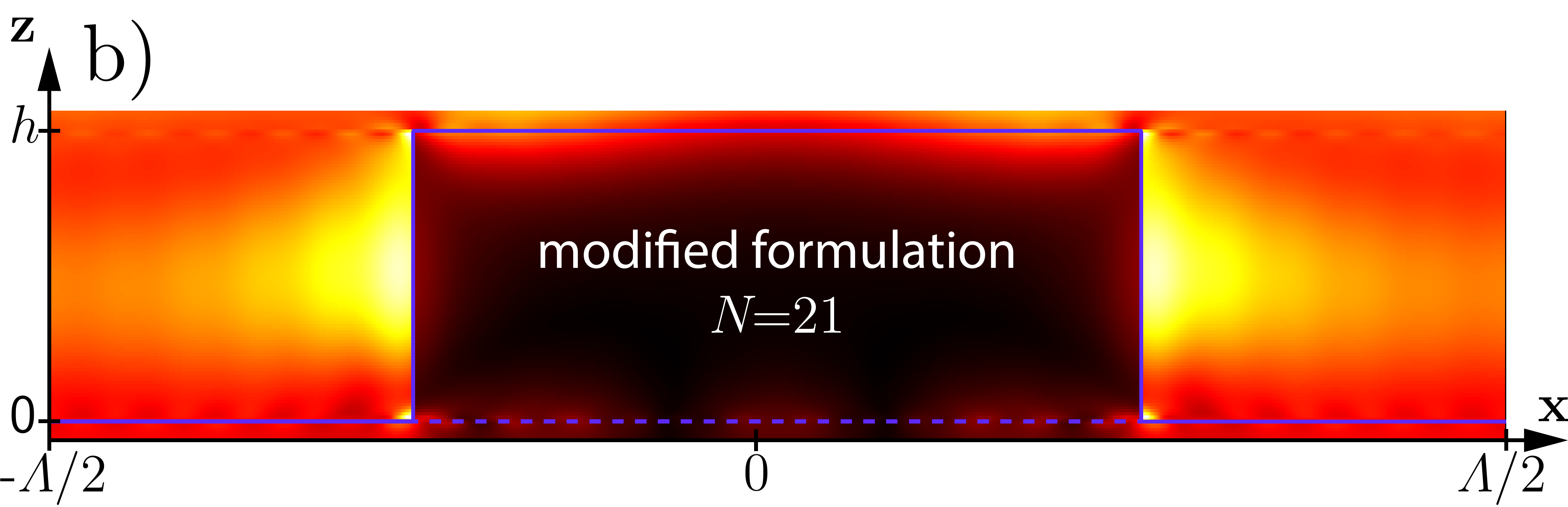}\\
\includegraphics[width=\linewidth]{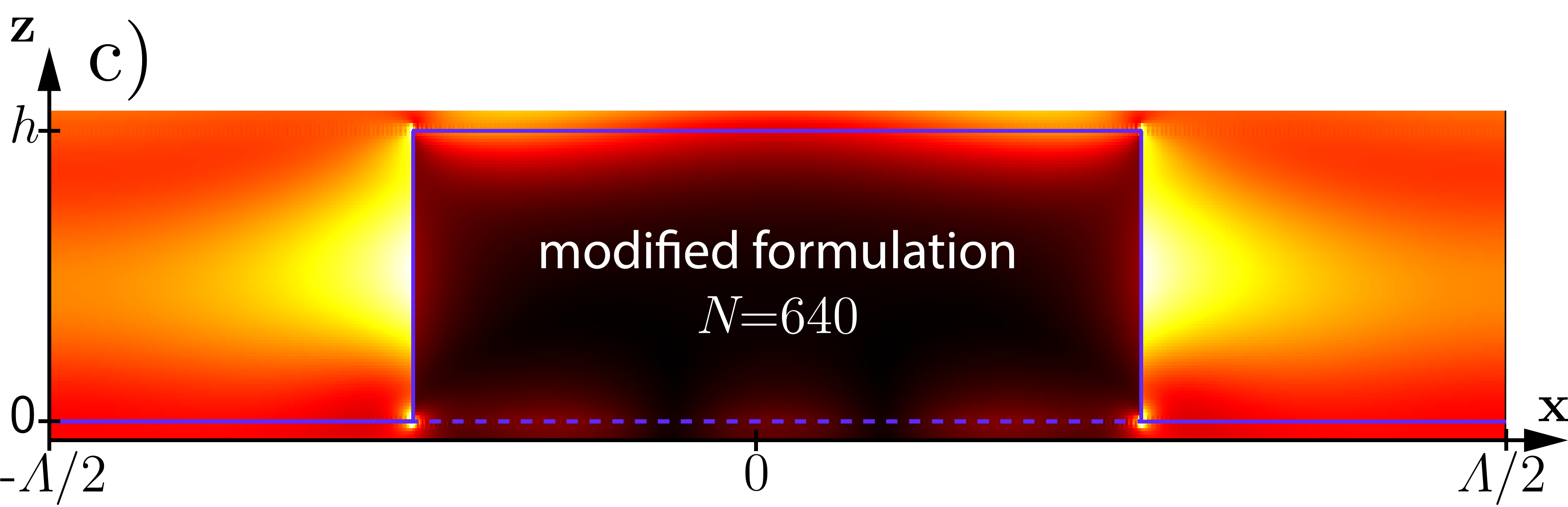}
}
\caption{Near-field distribution of the normal component of the electric field, $|E_n|=|E_x|$, in
and around the Au grating. a) Calculated with $N=21$ harmonics using the conventional RCWA method
(unphysical field oscillations can be observed). b) Calculated using the improved formulation,
$\tilde E_n=D_n/\varepsilon$, and $N=21$ harmonics. c) Calculated using the improved formulation
and $N=640$ harmonics.} \label{fig:FieldProfile1D}
\end{figure}

The spatial profile of the electric field in the grating region illustrates the full benefits of
the modified field calculation. \Figref{fig:FieldProfile1D}(a) depicts the conventional normal
field component $|E_n^{(21)}|=|E_x^{(21)}|$ in the gold grating for a moderately coarse
discretization of $N=21$ harmonics. It can be seen that $|E_n^{(21)}|$ exhibits unphysical
oscillations with a spatial frequency equal to the period of the smallest spatial frequency
component in the FS expansion of the solution. This is the well known Gibb's phenomenon, which
occurs when describing a discontinuous function with a truncated FS. On the other hand, the
modified normal field, $\tilde E_n$, does not suffer from such spurious oscillations at the
interface. In particular, even for a small number of harmonics, $N=21$, $\tilde E_n^{(21)}$ is
smooth, as per \figref{fig:FieldProfile1D}(b). At very large number of harmonics, $N=640$, the
modified normal field is free of any numerical artifacts, as can be seen in
\figref{fig:FieldProfile1D}(c).

\begin{figure}[t]
\centering \includegraphics[width=\linewidth]{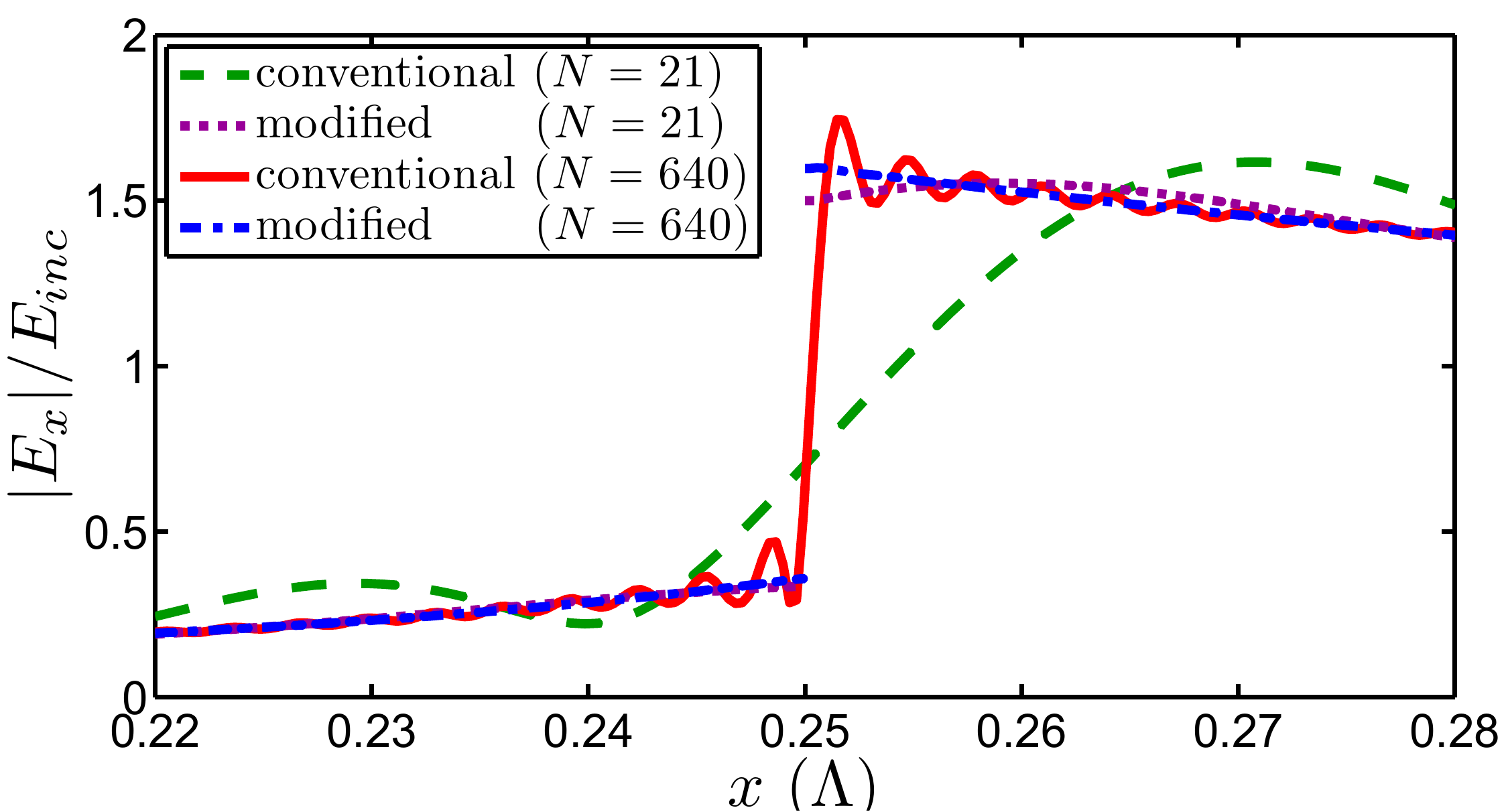} \caption{Closeup of the normal component of
the electric field, $|E_n|$ and $|\tilde E_n|$, near the metal-air interface at $z=h/2$, computed
using the conventional and modified RCWA methods, respectively.} \label{fig:SurfaceField1D}
\end{figure}

The improved formulation of RCWA exhibits another benefit, namely $\tilde E_n$ is by construction
discontinuous and exactly fulfills the corresponding boundary condition,
\begin{align}
    \varepsilon^\text{(in)} \E^\text{(in)}(\x_s)\cdot \n = \varepsilon^\text{(out)}\E^\text{(out)}(\x_s)\cdot \n, \label{eq:NormalBC}
\end{align}
at surface points of the grating, $\x_s$, where $\E^\text{in}$ and $\varepsilon^\text{(in)}$
($\E^\text{out}$ and $\varepsilon^\text{(out)}$) denote the electric field and permittivity inside
(outside) the grating, respectively, and $\n$ is the unit vector normal to the surface. In the
conventional formulation of the RCWA, the field $E_n^{(N)}$ does not satisfy \eqref{eq:NormalBC},
because $E_n^{(n)}$ is -- as a FS containing a \textit{finite} number of terms -- inherently
continuous. The closeup of the interfacial field around $x=0.25\Lambda$ in
\figref{fig:SurfaceField1D} emphasizes these ideas. Thus, the normal component of the field
calculated using the conventional formulation of RCWA shows spurious oscillations for both small
$N=21$ (dashed green) and high $N=640$ (solid red) number of harmonics, whereas the
modified formulation is free of such unphysical oscillations and discontinuous for any number of
harmonics, $N$ (see the dotted purple and dashed-dotted blue lines corresponding to $N=21$ and $N=640$,
respectively).

It should be clear now that the modified normal-field evaluation in 1D by means of the
displacement field represents an improvement of the conventional evaluation of $E_n$ in two ways:
First, it exhibits optimal self-convergence in the sense that it is as accurate as the far-field
and the continuous tangential field component. Second, it explicitly fulfills the boundary
condition \beqref{eq:NormalBC} at the material interfaces. Equally important, the improved
near-field calculation is achieved with minimal additional computational cost and does not alter
the mathematical framework of the core RCWA algorithm in 1D.

\section{Formulation of accurate near-field evaluation for 2D-periodic structures}
\label{sec:Formulation} The ideas of the previous section are straightforward to implement because
in 1D-periodic structures there is a trivial and unambiguous distinction between the tangential
and normal components of $\E$. However, this situation becomes more intricate in the case of
2D-periodic diffraction gratings.

As it will be shown now, accurate near-field evaluation is strongly related to correct Fourier
factorization in 2D-periodic structures. Fourier factorization means in this context the
decomposition of a product of periodic functions, $f=g\cdot h$, into its periodic factors $g$ and
$h$. Depending on the continuity properties of the factors and the product, different rules must
be applied in order to obtain fast convergence when increasing the number of FS terms. According
to these rules, if $f$ and $h$ possess simultaneous discontinuities, but $g$ is continuous at
those locations, the product rule yields fast convergence with respect to the number of FS terms,
namely one uses $f = g\cdot h$. Moreover, if $g$ and $h$ are simultaneously discontinuous, but $f$
is continuous, the inverse rule should be used, i.e., $f$ must be factorized as $f=(1/g)^{-1}
\cdot h$. A rigorous explanation of these rules and the solution to the 1D Fourier factorization
problem is given in \cite{Li1996josaa}.

Because of the trivial distinction between the continuous (tangential) and discontinuous (normal)
components of the electric field in 1D, Fourier factorization is straightforward in that case. But
for 2D-periodicity, three different approaches to achieve the correct Fourier factorization have
been proposed: {\itshape i)} Approximate the material boundaries by a coordinates-aligned
staircase-contour \cite{l97josaa}. {\itshape ii)} Devise a coordinate system in which a given
grating geometry is coordinate system aligned and use approach {\itshape i} to obtain the correct
Fourier factorization \cite{weis09oe,eb10oe}. {\itshape iii)} Construct a normal vector field
(NVF) to decompose $\E$ into its normal and tangential components and then apply the corresponding
correct factorization rules to them \cite{srk07josaa}.

Since the accurate near-field evaluation relies on the decomposition of the electric field into
normal and tangential components, it is natural to use the factorization approach {\itshape
(iii)}, the NVF approach. It is out of the scope of this paper to derive the full formulation of
2D-RCWA, so that only the crucial and unconventional steps will be given here. Thus, the normal
and tangential components of $\D=\varepsilon \E$ have to be decomposed using the product- and
inverse rule, respectively. This leads to the following relations for the $\alpha$-component of
the displacement field \cite{pn01josaa,srk07josaa,gt10josaa}:
\begin{align}
\fvec{D_\alpha} &= \varepsilon_0\sum_{\beta=1}^3 \left(\delta_{\alpha,\beta}\fmatrix\varepsilon -
\Delta_{\alpha\beta}\right)\fvec{E_\beta}, \label{eq:ConstRelationNVF}
\end{align}
where $\delta_{\alpha\beta}$ is the Kronecker delta. Here, $\fvec{f}$ denotes the vector of FS
coefficients of a scalar function $f$, $\fmatrix{g}$ is the Toeplitz matrix of FS coefficients of
$g$, and the matrix $\Delta_{\alpha\beta}$ is given by $\Delta_{\alpha\beta} = \frac
12\left(\Delta \fmatrix{N_\alpha N_\beta} +\fmatrix{N_\alpha N_\beta}\Delta\right)$ with $\Delta =
\fmatrix{\varepsilon} - \fmatrix{1/\varepsilon}^{-1}$ and $N_\alpha$ is the $\alpha$-component of
the NVF, $\NVF=(N_1,N_2,N_3)^T$, of the material boundary. The matrix
$\delta_{\alpha,\beta}\fmatrix\varepsilon- \Delta_{\alpha\beta}$ implements the three steps of the
Fourier factorization: the decomposition of $\E$ into normal and tangential components ($N_\beta$
in $\fmatrix{N_\alpha N_\beta}$), factorization using either the inverse rule
($\fmatrix{1/\varepsilon}^{-1}$) or the product rule ($\fmatrix{\varepsilon}$), and
back-projection to Cartesian coordinates ($N_\alpha$ in $\fmatrix{N_\alpha N_\beta}$).

Note that one can also use asymptotically equivalent definitions of $\Delta_{\alpha\beta}$.
However, our investigations have shown that despite the fact that the choices
$\Delta_{\alpha\beta} = \Delta \fmatrix{N_\alpha N_\beta}$ and $\Delta_{\alpha\beta} =
\fmatrix{N_\alpha N_\beta} \Delta$ produce similar convergence speed, they do not conserve the
power for lossless structures, whereas the choice $\Delta_{\alpha\beta} =\fmatrix{N_\alpha} \Delta
\fmatrix{N_\beta}$ yields power conservation but at the price of slower convergence. All three
formulations are asymptotically equivalent with respect to the number of terms in the FS
expansion, due to the commutation of Toeplitz operators \cite{pn01josaa}.

Normal vector fields can be constructed analytically for a variety of structures and automated
algorithms to obtain a NVF for arbitrary grating geometries have been developed \cite{gotz08oe}.
It should be noted that this formulation allows inclined NVFs, i.e. NVFs with simultaneously
non-vanishing $x$, $y$, and $z$ components. Such a NVF is required to accurately model obliquely
etched structures (see \figref{fig:Structures}(d)) and hence can be viewed as a generalization
of the methods presented in \cite{gt10josaa} ($N_y=0$, \figref{fig:Structures}(b)) and
\cite{srk07josaa} ($N_z=0$, \figref{fig:Structures}(c)).

Given the constitutive relation \beqref{eq:ConstRelationNVF}, one can derive the RCWA
eigenvalue-problem
\begin{align}
\textsf{M} \left(\begin{array}{c} \fvec{S_x}\\\fvec{S_y}\\\fvec{U_x}\\\fvec{U_y}
\end{array}\right)=\beta \left(\begin{array}{c}
\fvec{S_x}\\\fvec{S_y}\\\fvec{U_x}\\\fvec{U_y}
\end{array}\right) \label{eq:defEVproblem}
\end{align}
for the electromagnetic modes described by $\fvec{S}$ and $\fvec{U}$ and propagation constant,
$\beta$. The most general formulation of the system-matrix $\textsf{M}$ reads:
{\footnotesize
\begin{align*}
\!\textsf{M}\!=\!\!\left(\!\!\!\!\begin{array}{cccc}
 \textsf{K}_x \textsf{B}\Delta_{zx} & \textsf{K}_x \textsf{B}\Delta_{zy} \\
 \textsf{K}_y \textsf{B}\Delta_{zx} & \textsf{K}_y \textsf{B}\Delta_{zy} \\
 \Delta_{yx} + \Delta_{yz}\textsf{B}\Delta_{zx}- \textsf{K}_x \textsf{K}_y \!\! & \!\!\textsf{K}_x \textsf{K}_x + \Delta_{yz} \textsf{B}\Delta_{zy}- \textsf{C}_y \!\! \\
 \textsf{C}_x - \Delta_{xz}\textsf{B}\Delta_{zx} - \textsf{K}_y \textsf{K}_y \!\! & \!\!\textsf{K}_y \textsf{K}_x-\Delta_{xz} \textsf{B}\Delta_{zy}- \Delta_{xy} \!\!
\end{array}\right. \\
\left.\!\!\!\!\begin{array}{cccc}
 \textsf{K}_x \textsf{B}\textsf{K}_y &\textsf{I}- \textsf{K}_x \textsf{B}\textsf{K}_x \\
 \textsf{K}_y \textsf{B}\textsf{K}_y - \textsf{I} & - \textsf{K}_y\textsf{K}_x\\
 \Delta_{yz} \textsf{B}\textsf{K}_y & - \Delta_{yz} B\textsf{K}_x\\
-\Delta_{xz} \textsf{B}\textsf{K}_y & \Delta_{xz} \textsf{B}\textsf{K}_x
\end{array}\!\!\!\!\right),
\end{align*} }
with $\textsf{B}=\left(\fmatrix{\varepsilon}-\Delta_{zz}\right)^{-1}$ and
$\textsf{C}_\alpha=\fmatrix{\varepsilon}-\Delta_{\alpha\alpha}$. The matrices
$\textsf{K}_\alpha=\operatorname{diag}\left(k_{\alpha n}\right)$ are diagonal matrices of the
in-plane propagation constants, $k_{xn}$ for $\alpha=x$ and $k_{yn}$ for $\alpha=y$, of the
diffraction orders and $\textsf{I}$ is the identity matrix. The size of matrices $\textsf{B}$,
$\textsf{C}_\alpha$, $\textsf{K}_\alpha$, and $\textsf{I}$ is $N_0\times N_0$, whereas the size of
$\textsf{M}$ is $4N_0\times4N_0$, where $N_0=(2N+1)^2$ is the total number of FS coefficients in
the calculation and the same number of FS terms for the truncation of the FS in the $x$- and
$y$-direction is assumed.

The mode amplitudes $\fvec{S}$ and $\fvec{U}$ are determined by matching the tangential components
of the electromagnetic field at the top and bottom of the grating region, whereas multilayered
structures are described by the staircase approximation in conjunction with the numerically stable
$\mathcal{S}$-matrix algorithm. This yields the FS coefficients, $\fvec{\E}$, of the electric
field everywhere in the grating region, in the cover, and the substrate.

Let us now denote by $\rec{\fvec{f}}$ the FS reconstruction of a Fourier coefficient
vector $\fvec{f}$,
\begin{align*}
  \rec{\fvec{f}}(x,y) = \sum_{n=-N}^{N}\sum_{m=-N}^{N} f_{nm} \exp\left(i m \frac{2\pi}{\Lambda_1}x +i n
  \frac{2\pi}{\Lambda_2}y\right),
\end{align*}
where $f_{nm} = \fvec{f}_{ (n+N)(2N+1)+m+N+1 }$ are the $N_0$ FS coefficients of $f$. Within the
conventional RCWA framework, each component of $\E$ is evaluated as:
\begin{align}
E_\alpha^{(N)} = \rec{\fvec{E_\alpha}}. \label{eq:conventional}
\end{align}
This relation does not take into account the continuity properties of the different components of
$\E$ and hence will lead to spurious oscillations and slow convergence of the near-field as it was
seen in section~\ref{sec:1D}.

The modified field evaluation for 2D-periodic diffraction gratings requires one to define the
continuous normal component of $\D$ and the tangential component of $\E$. Their FS coefficient
vectors are given by:
\begin{subequations}
\begin{align}
  \fvec{\D_n} &= \frac 12 \varepsilon_0 \left(\fmatrix{\NVF \NVF^T}\fmatrix{1/\varepsilon}^{-1} + \fmatrix{1/\varepsilon}^{-1} \fmatrix{\NVF \NVF^T}\right)\fvec{\E},\\
  \fvec{\E_t} &= \fmatrix{\idmat - \NVF \NVF^T}\fvec{\E},
\end{align}
\end{subequations}
where $\idmat$ denotes the $3\times 3$ identity matrix and $\NVF \NVF^T$ is the $3\times 3$
projection matrix defined by the NVF at any point in space, $\x$. Note that by construction,
$\fvec{\D_n}$ and $\fvec{\E_t}$ are FS coefficient vectors of vector fields that are continuous at
material interfaces. Hence their reconstructions, $\rec{\fvec{\D_n}}$ and $\rec{\fvec{\E_t}}$, do
not suffer from Gibb's phenomenon at the interface. With this observation in mind, the electric
field in the improved RCWA method for 2D-periodic structures at a point, $\x$, is given by:
\begin{align}
  \tilde \E^{(N)}(\x) = \varepsilon_0^{-1}\rec{\fvec{\D_n}}(\x) / \varepsilon(\x) +  \rec{\fvec{\E_t}}(\x) \label{eq:accurateGeneric}
\end{align}
and is expected to yield fast near-field convergence and non-oscillatory spatial field profiles.
To investigate the validity of these predictions, the accurate field evaluation was implemented in
a commercially available RCWA computer software, OmniSim/RCWA \cite{omnircwaPD}.


It should be noted that the other electromagnetic fields can be easily calculated with our
improved method, too. Specifically, the displacement field, $\D$, can be evaluated using the
modified electric field $\tilde E$, namely $\D^{(N)}(\x) = \varepsilon_0\varepsilon(\x)\tilde
\E^{(N)}(\x)$, and hence will have the same convergence properties as $\tilde \E$. The magnetic
induction $\Bfield$ and the magnetic field $\Hfield=\mu\Bfield$ do not require special attention,
because they are continuous in non-magnetic materials and hence behave similar to the continuous
tangential component of the electric field.

\section{Quantification the accurate near-field evaluation in 2D-periodic structures}
\label{sec:2D} In this section, the improved formulation for accurate near-field calculations in
2D-periodic structures is assessed using two test structures under different configurations. To
this end, we first extend the definition of the grating norm \beqref{eq:NormG} of a scalar or
vector function, $f$, to 2D-periodic structures in the following straightforward way:
\begin{align}
\| f \|_G  &= \left(\int_{0}^{h}\int_{-\Lambda_2/2}^{\Lambda_2/2}
\int_{-\Lambda_1/2}^{\Lambda_1/2} \vert f(x,y,z)\vert^2\ dx\ dy\ dz \right)^{\nicefrac{1}{2}},
\label{eq:NormG2D}
\end{align}
where the integral is evaluated over the three-dimensional grating region.



\subsection{Analysis of a 1D-periodic grating using 2D-RCWA}
\begin{figure}[t]
\centering
\minibox{\includegraphics[width=\linewidth]{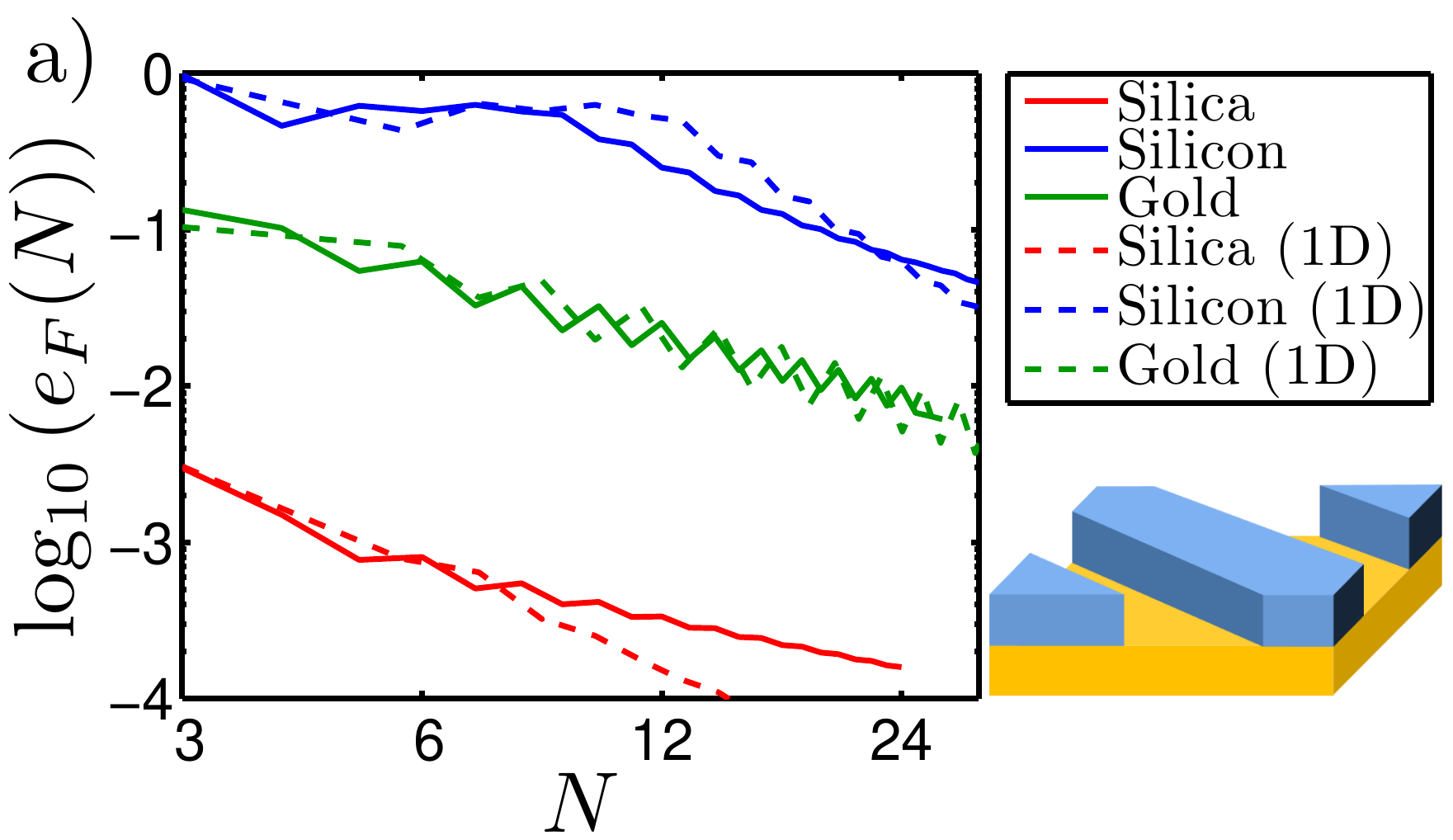}\\
\includegraphics[width=\linewidth]{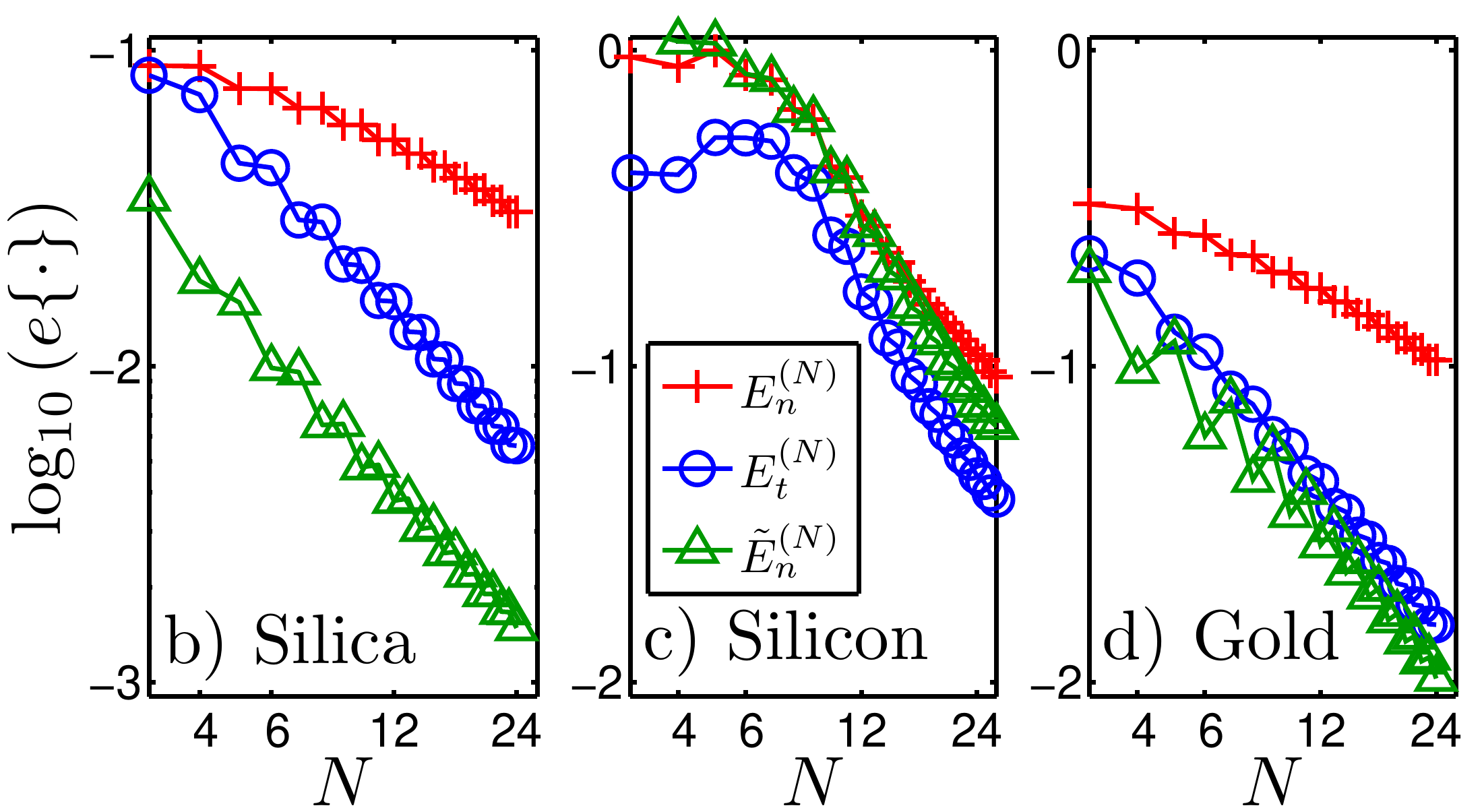}}
\caption{Computational results for a rotated binary 1D grating. a) Error of calculated far-field
vs. $N$, determined for three gratings made of different materials. The decrease in the error
follows that of the far-field (dashed lines) of the 1D simulations (see also
section~\ref{sec:1D}). b), c), d) Near-field error corresponding to the silica, silicon, and gold
grating, respectively.} \label{fig:15DOverview}
\end{figure}

The first 2D-periodic diffraction grating under consideration is a 1D binary grating, as shown in
\figref{fig:Structures}(a), rotated by $\pi/4$ in the $x-y$-plane. It should be obvious that it
can be modeled as a double-periodic $2D$ grating with periods
$\Lambda_1=\Lambda_2=\sqrt{2}\tilde\Lambda=\SI[parse-numbers = false]{\sqrt{2}}{\micro\metre}$,
where $\tilde\Lambda=\Lambda_1$ is the period of the grating when it is viewed as a 1D-periodic
structure. For clarity, the primary unit cell of the grating is depicted in the inset of
\figref{fig:15DOverview}(a). With this choice, the reference quantities $R^{ref}, T^{ref}$, and
$\E^{ref}$ can be calculated using 1D simulations, an approach inspired by an example in
\cite{srk07josaa}.

The error in the calculation of the far-field, $\errsym_F$ from \eqref{eq:defFarFieldError}, when
2D simulations are employed is depicted in \figref{fig:15DOverview}(a). The convergence of the
calculations for the silicon and gold gratings closely follows the convergence trends observed
when 1D simulations are performed and, as expected, it shows somewhat worse, yet still good,
agreement for the silica structure. This difference is explained by the fact that the NVF
introduced in the 2D-RCWA formulation is discontinuous away from material interfaces and hence it
can degrade the convergence rate, especially for the low-index of refraction contrast case.

\begin{figure}[htbp]
\centering \includegraphics[width=\linewidth]{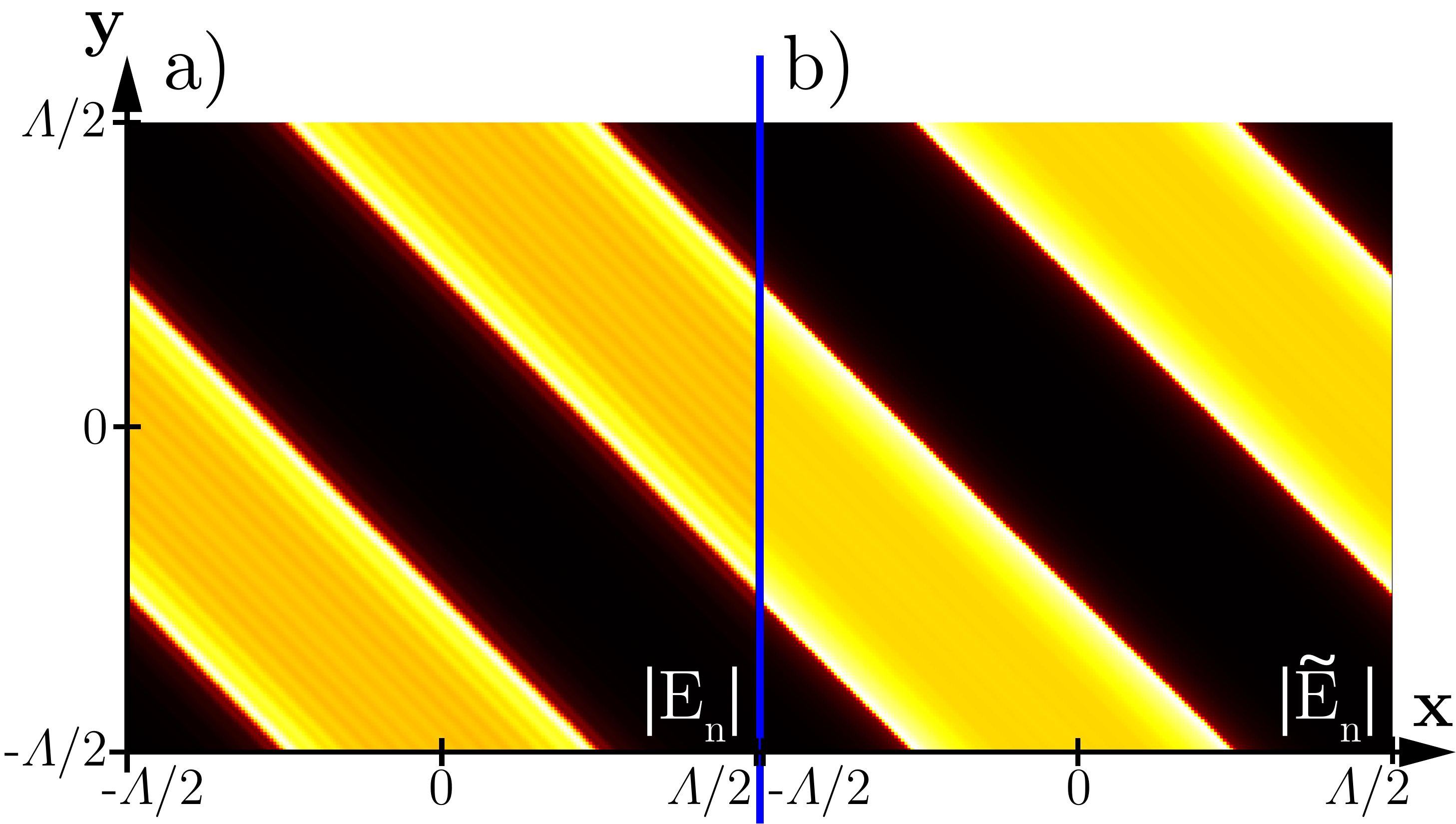} \caption{a) Normal component of the
electric field, $|E_n|$, in the grating region at $z=h/2$, determined by using the conventional
RCWA and $N=27$. b) Normal electric field component, $|\tilde E_n|$, determined for the same
grating parameters as in a) but using the improved algorithm. The blue vertical line was added for
clarity and merely separates the two plots.} \label{fig:FieldProfile15D}
\end{figure}
The conclusions of our analysis of the convergence of the near-field are summarized in
figures~\ref{fig:15DOverview}(b)--\ref{fig:15DOverview}(d). Thus, for both the silica and gold
diffraction gratings, the normal component of the modified electric field, $\tilde \E$, exhibits
faster convergence than this same field component calculated using the conventional form of RCWA.
In both cases, $\Errsym{\tilde E_n^{(20)}}$ is one order of magnitude smaller than
$\Errsym{E_n^{(20)}}$. For the silicon grating, only marginal differences between the two
formulations can be observed. This is in agreement with the results obtained in the 1D case, as
per \figref{fig:SelfConvergenceField1D}(b). For small $N$, $\tilde \E^{(N)}$ and $\E^{(N)}$ are
determined with a comparable degree of accuracy, but for the larger number of harmonics considered
in \figref{fig:15DOverview}(c), i.e. $N=30$, a higher accuracy of our improved formulation of
the RCWA can clearly be observed.

These conclusions are further validated by the profile of the electric field, as presented in
\figref{fig:FieldProfile15D}. This figure shows the spatial distribution of the normal component
of the electric fields, $\E^{(27)}$ and $\tilde \E^{(27)}$, calculated in the median plane of the
grating. A simple examination of these field profiles confirms that the spurious oscillations of
the field $\tilde\E^{(27)}$ near the surface have much smaller amplitude as compared to that of
the variations of $\E^{(27)}$. Moreover, a closer inspection of the surface-fields shows that the
boundary condition \beqref{eq:NormalBC} is fulfilled by $\tilde \E^{(27)}$ only. This first
test-case already reveals that the improved near-field evaluation is more accurate in the case of
2D-periodic structures, too.

\subsection{Near-field calculations for an intrinsically 2D-periodic grating}
In order to thoroughly test the near-field evaluation for 2D-periodic structures using the
improved RCWA presented in this article, in what follows we consider the challenging test
structure depicted in \figref{fig:Structures}(c). The grating region consists of a coordinate
system aligned parallelepiped with the length of the sides aligned to the $x$-, $y$-, and $z$-axis
being $a=0.5\Lambda_1$, $2a$, and $h$, respectively, placed adjacently to a semicircular cylinder
with radius $a$ and height $h$, with $\Lambda=\Lambda_1=\Lambda_2=\SI{0.25}{\micro \meter}$. The
structure is illuminated normally by a $x$-polarized plane wave with wavelength
$\lambda=\SI{0.5}{\micro \meter}$.

Since it is generally computationally time consuming to obtain high-accuracy solutions in the case
of 2D-periodic structures and due to the fact that the higher the ratio $\Lambda/\lambda$ and the
refractive index $|n|$, the more harmonics are necessary to achieve convergence
\cite{popov2002staircase}, a relatively small period-to-wavelength ratio of
$\Lambda/\lambda=\SI{0.25}{\micro \meter}/\SI{0.5}{\micro \meter}=0.5$ is chosen for this example.

\begin{figure}[t]
\centering \includegraphics[width=\linewidth]{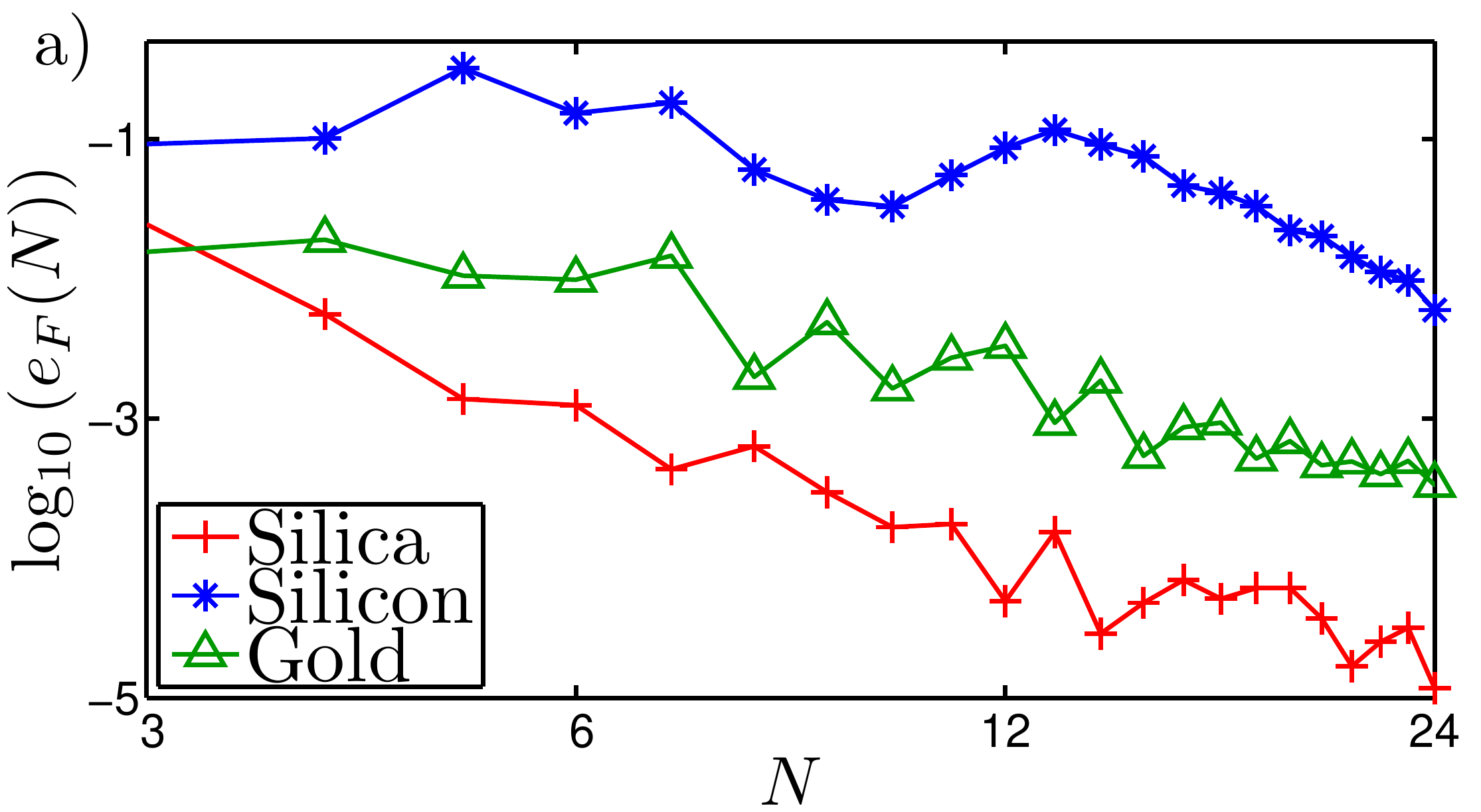}
\includegraphics[width=\linewidth]{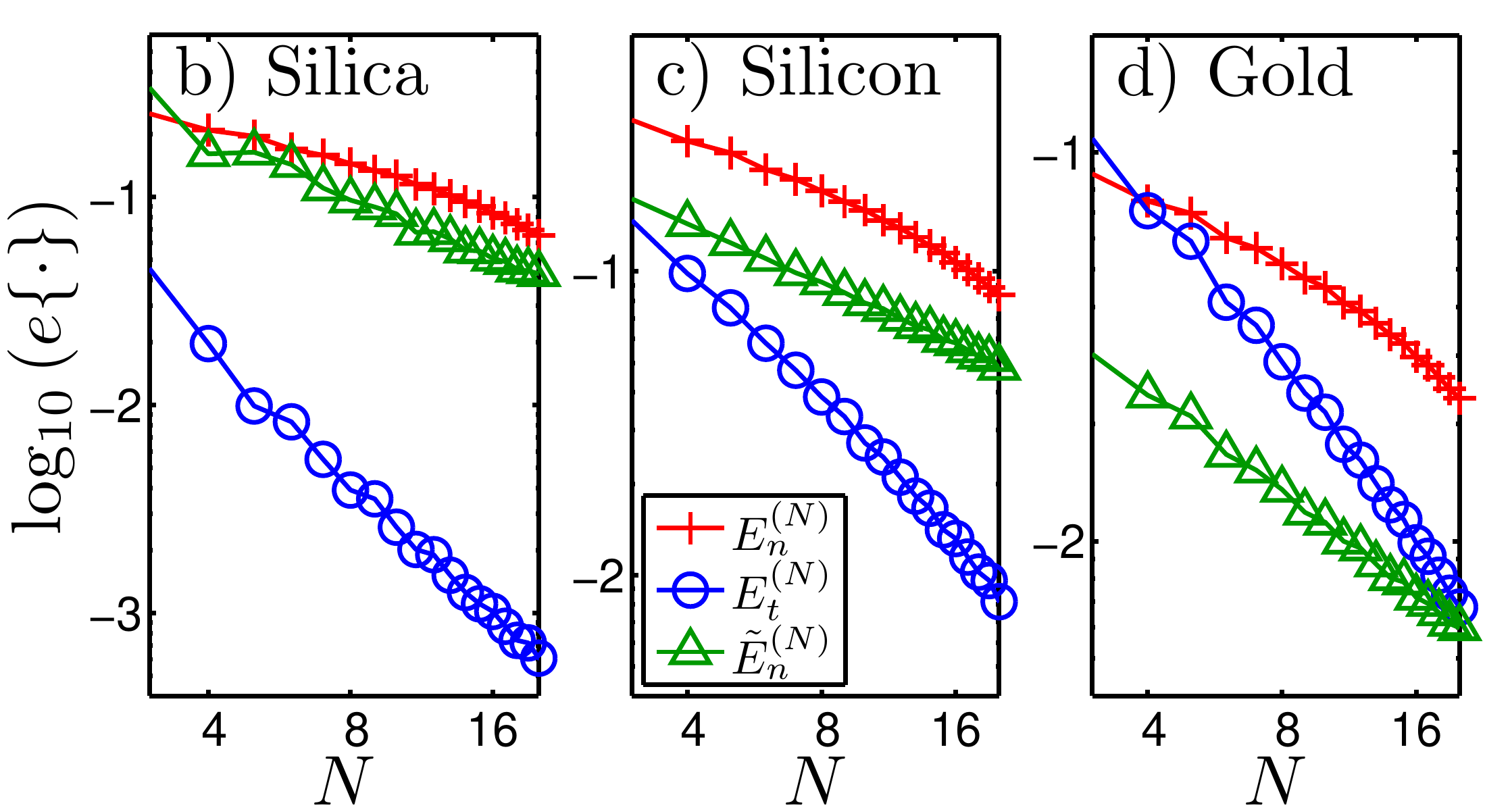} \caption{Computational results
for the 2D grating shown in \figref{fig:Structures}(c). a) Error of calculated far-field vs.
$N$, determined for three gratings made of different materials. b), c), d) Near-field error
corresponding to the silica, silicon, and gold grating, respectively.} \label{fig:2DOverview}
\end{figure}

As reference values in the definition of the far-field error, $\errsym_F$ from
\eqref{eq:defFarFieldError}, $T^{ref}=T^{(31)}$ and $R^{ref}=R^{(31)}$, namely results obtained
from simulations with $N=31$ are chosen. The convergence of the far-field, $\errsym_F(N)$, is
shown in \figref{fig:2DOverview}(a), where the far-field physical quantity considered are the
transmission and reflection coefficients, $T$ and $R$, respectively. As expected, the fastest
convergence can be observed for the silica grating, because it has a low refractive index. The
numerical error obtained for the gratings made of gold and silicon are one and two orders of
magnitude larger than in the case of the silica grating, respectively. This behavior is similar to
that seen in the 1D case for a small number of harmonics, $N<30$ (cf.
\figref{fig:SelfConvergence1D}).

Figures~\ref{fig:2DOverview}(b)--\ref{fig:2DOverview}(d) contain the dependence of the near-field
self-error determined for the three material configurations, silica, silicon, and gold,
respectively. The three physical quantities plotted in each case are the $z$-component of $\E$ and
the in-plane components, $\E_{xy} := (E_x,E_y,0)^T$ and $\tilde \E_{xy} := (\tilde E_x,\tilde
E_y,0)^T$, obtained by using the conventional and improved versions of RCWA, respectively. Note
that the in-plane component contains the discontinuous normal component, $\NVF\cdot\E$
($\NVF\cdot\tilde\E$), and the continuous tangential component, $(\idmat - \NVF\NVF^T)\E_{xy}$
($(\idmat - \NVF\NVF^T)\tilde \E_{xy}$).

It can be seen that in all cases, the $z$-component of $\E$, which is continuous at vertical
surfaces inside the grating region, converges much faster than the in-plane component, in both the
conventional and modified formulations. In addition, the modified formulation leads to a somewhat
smaller error than the conventional formulation. Specifically, it was found that $\Errsym{\tilde
\E_{xy}^{(N)}}\approx 0.5 \Errsym{\E_{xy}^{(N)}}$ for the silica grating and $\Errsym{ \tilde
\E_{xy}^{(N)}}\approx 0.9 \Errsym{\E_{xy}^{(N)}}$ for the silicon and gold gratings. However, the
corresponding convergence speed, i.e. the slope of $\Errsym{\E_{xy}^{(N)}}$ and $\Errsym{\tilde
\E_{xy}^{(N)}}$, is the same. Moreover, in the modified formulation of the RCWA the convergence
speed of the tangential component of the near-field is larger than that of the in-plane component.
Three factors contribute to this behavior: {\itshape i)} The decomposition of the near-field in
a normal and tangential component in 2D-periodic structures relies on the specific definition of
the normal vector field, $\NVF$ and it is not directly performed in Cartesian coordinates as in
the 1D case. Hence, the inexact field decomposition by $\NVF$ introduces an additional error.
{\itshape ii)} The field of normal vectors characterizing the structure is only uniquely defined
at the interfaces defining the grating, except at the corners, and, more importantly, away from
the grating surface. This ambiguity allows and can lead to choices of NVFs, which are not optimal
for the convergence and accurate calculation of the near-field. {\itshape iii)} The normal vector
field itself has discontinuities, which can cause additional oscillations in the spatial profile
of the electromagnetic field.

\begin{figure}[t]
\centering \includegraphics[width=\linewidth]{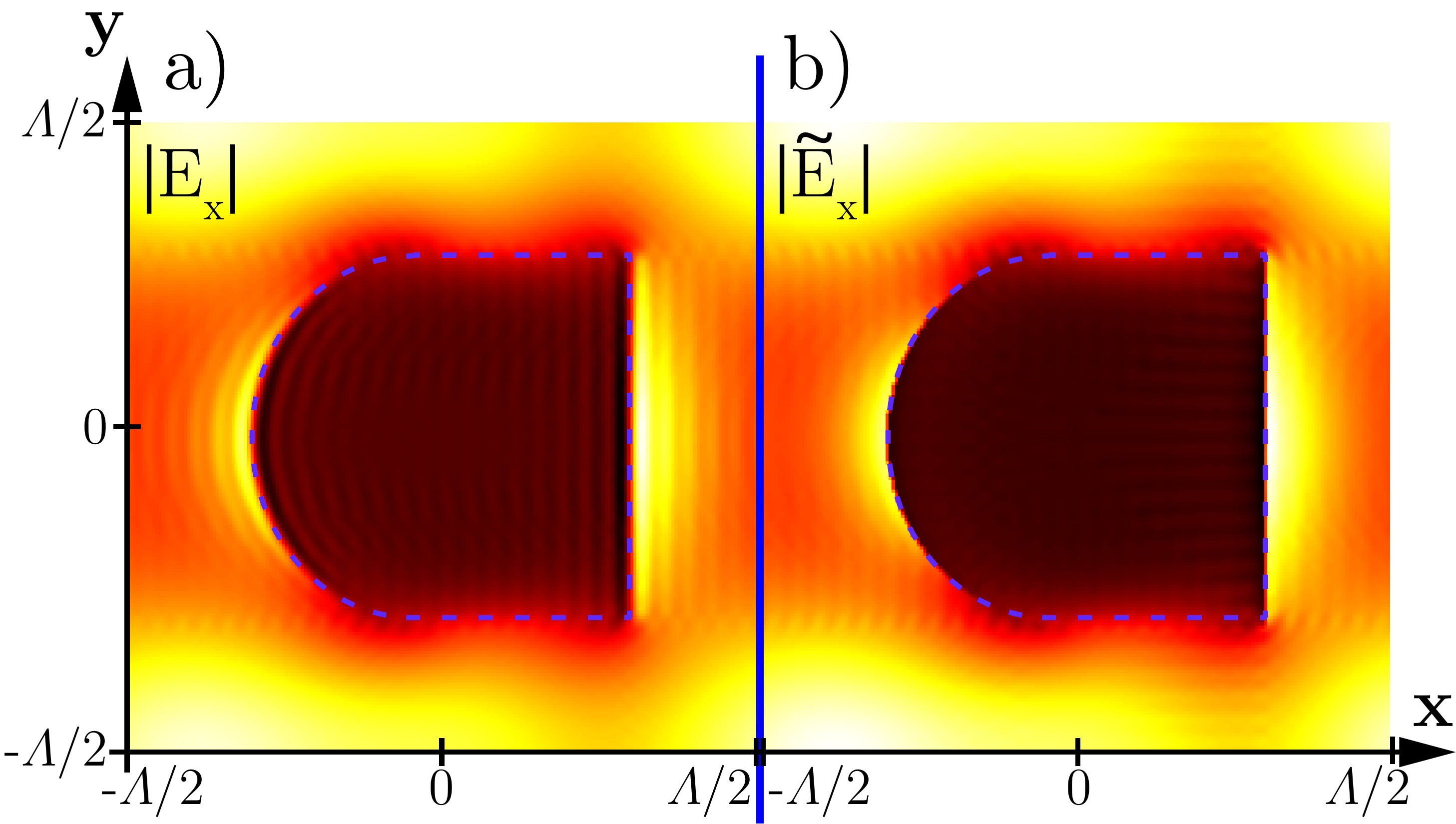} \caption{a) Spatial distribution of the
dominant component of the electric field, $|E_x|$, in the 2D grating region at $z=h/2$, determined
by using the conventional RCWA and $N=27$. b) Spatial distribution of the dominant component of
the electric near-field, $|\tilde E_x|$, determined for the same grating parameters as in a) but
using the improved algorithm. The blue vertical line was added for clarity and merely separates
the two plots.} \label{fig:FieldProfile2D}
\end{figure}

It is also worthwhile to investigate the spatial profile of the near-field. The dominant
$x$-component of the electric field in a horizontal cross-section through the grating region at
$z=h/2$ is depicted in \figref{fig:FieldProfile2D}. As in the 1D case, these maps show that the
field $\tilde E_x$ exhibits spatial oscillations with smaller amplitude as compared to the
variations of the field $E_x$, especially at $y$-aligned interfaces (outlined with blue dashed
lines in \figref{fig:FieldProfile2D}).

\section{Out-of-plane normal vector fields for oblique diffraction gratings}
\label{sec:1D2Dz} Oblique diffraction gratings as those shown in figures~\ref{fig:Structures}(b) and
\ref{fig:Structures}(d) are modeled in the RCWA within the staircase approximation along a
direction perpendicular onto the grating plane. Each computational slice is assumed to be a
$z$-independent structure, for which the eigenmodes can be found numerically by solving
\eqref{eq:defEVproblem}. The field in each computational layer is then found by using the boundary
conditions at the top and bottom interfaces of the grating, employing a numerically stable
$\mathcal{S}$-matrix formulation. The validity of this staircase approximation for 1D-periodic
gratings under TE-polarization has been proven in \cite{popov2002staircase}. In the context of the
NVF formulation of RCWA for oblique 1D-periodic structures, it has been found that the use of an
out-of-plane NVF, i.e. $\NVF = \left(N_x,0,N_z\right)^T$, is beneficial for the far-field
convergence speed \cite{popov2002staircase, gt10josaa}. In this section we will study the relation
between the accurate field formulation~\beqref{eq:accurateGeneric} and the near-field convergence
speed for slanted 1D-periodic structures.

The situation for oblique 2D-periodic structures has not yet been explored in the context of RCWA.
However, based on the results presented so far, one can conjecture that both the near- and
far-field convergence of RCWA can be improved by using an out-of-plane 3D NVF, $\NVF =
\left(N_x,N_y,N_z\right)^T \neq 0$, and that the near-field evaluation would be more accurate as
well. The validity of this supposition will be explored in the second part of this section.

\subsection{Analysis of slanted 1D-periodic binary diffraction gratings}
\label{sec:sub1Dz} In order to analyze oblique 1D-periodic structures, we consider the slanted
binary grating depicted in \figref{fig:Structures}(b). The period of the grating is
$\Lambda=\SI{1}{\micro \meter}$, the filling factor, $\rho=0.5$, the height,
$h=\SI{0.25}{\micro\meter}$, and the slanting angle is $\theta=\pi/4$. Only the gold grating is
considered in this section as this would be the most challenging case. If the unit cell is assumed
to extend from $x=-\Lambda/2$ to $x=\Lambda/2$ and the center of the binary grating is set to be
$x=0,~z=h/2$, a suitable out-of-plane NVF is given by $\tilde\NVF(x,y,z) =
\operatorname{sign}\left(x-z\right)\sqrt{2}\left(1,0,1\right)^T$.
\begin{figure}[t]
\centering
\minibox{\includegraphics[width=\linewidth]{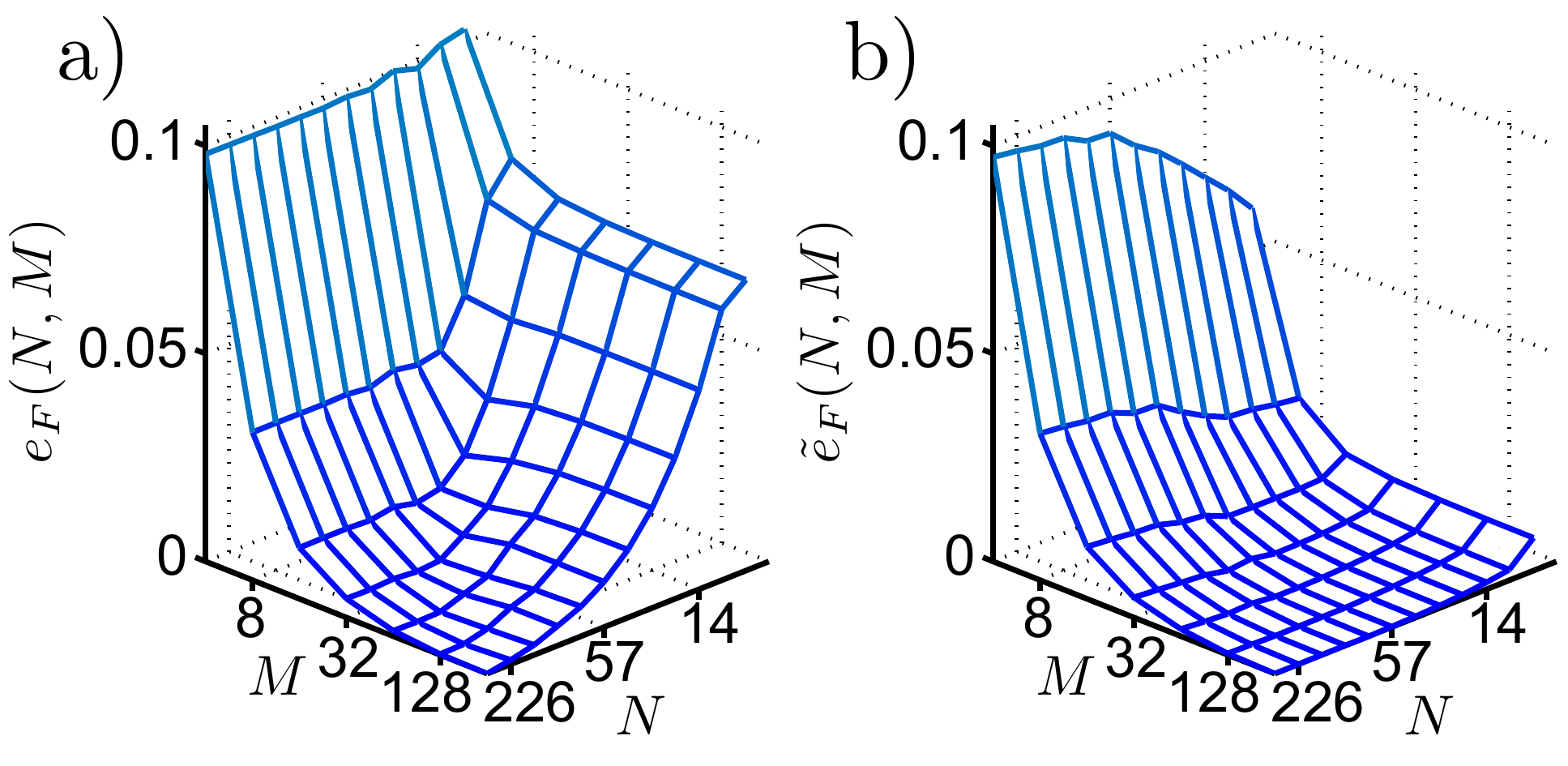} \\
\includegraphics[width=\linewidth]{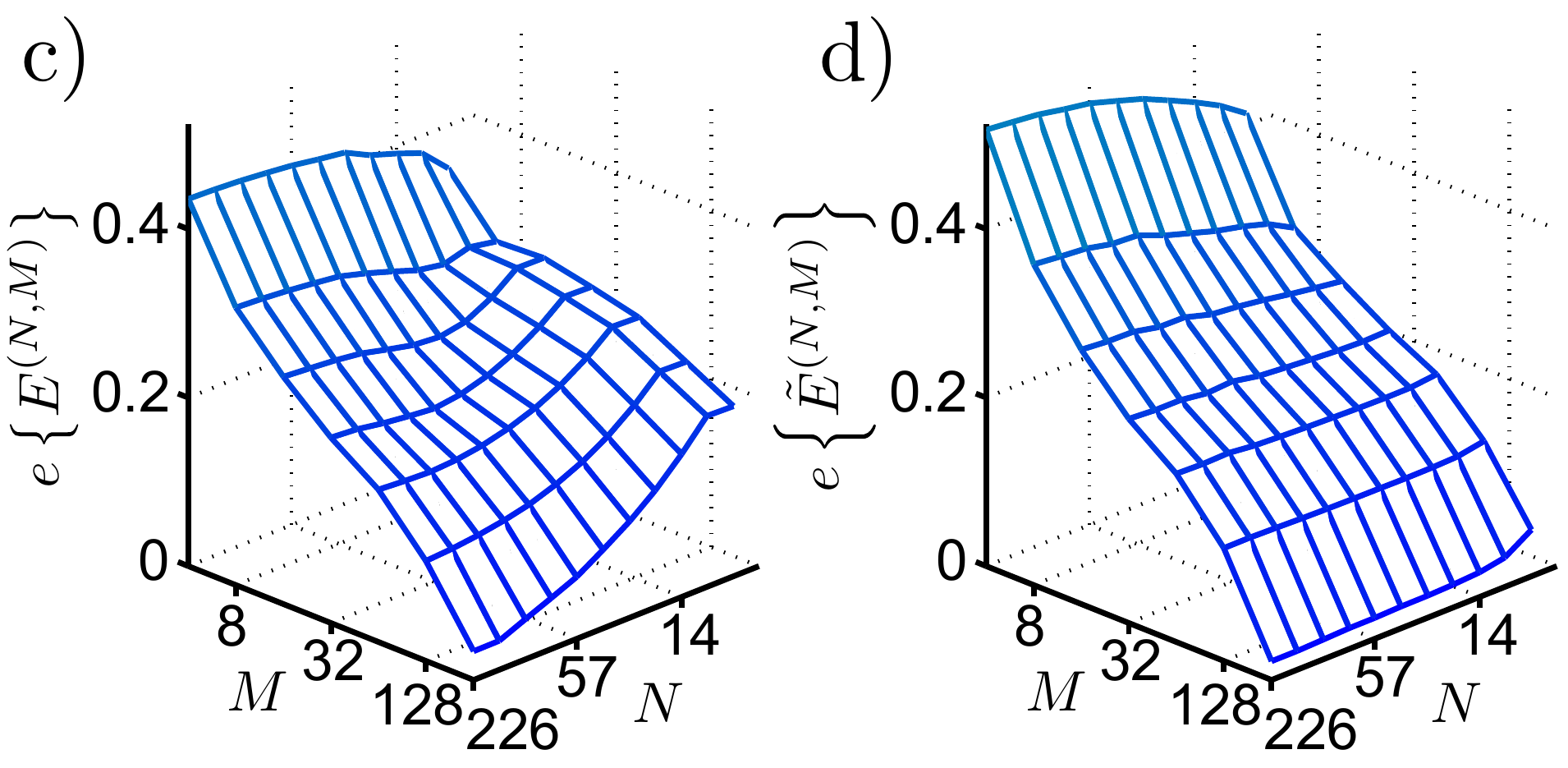} }
\caption{Computational results for the slanted 1D binary grating shown in
\figref{fig:Structures}(b). a) Far-field self-error $\errsym_F(N,M)$ vs. $N$ and $M$
corresponding to the in-plane NVF formulation. b) Far-field self-error $\tilde \errsym_F(N,M)$
corresponding to the out-of-plane NVF formulation. c) Near-field self convergence
$\Errsym{\E^{(N,M)}}$ corresponding to in-plane NVF. d) Near-field self convergence
$\Errsym{\tilde \E^{(N,M)}}$ corresponding to out-of-plane NVF.} \label{fig:1DzOverview}
\end{figure}

The two formulations of the RCWA compared in this section are the in-plane NVF, which is used in
conventional RCWA together with the conventional field evaluation \beqref{eq:conventional}, and the
out-of-plane NVF, $\tilde\NVF$, combined with the improved field evaluation formulation
\beqref{eq:accurateGeneric}. In contrast to the results presented in the previous sections, the two
formulations yield different results for both the near- and far-field quantities. Moreover, for the sake
of the clarity of the presentation, all physical quantities corresponding to the out-of-plane NVF
formulation are denoted with a tilde symbol.

\begin{figure}[t]
\centering \minibox{
\includegraphics[width=\linewidth]{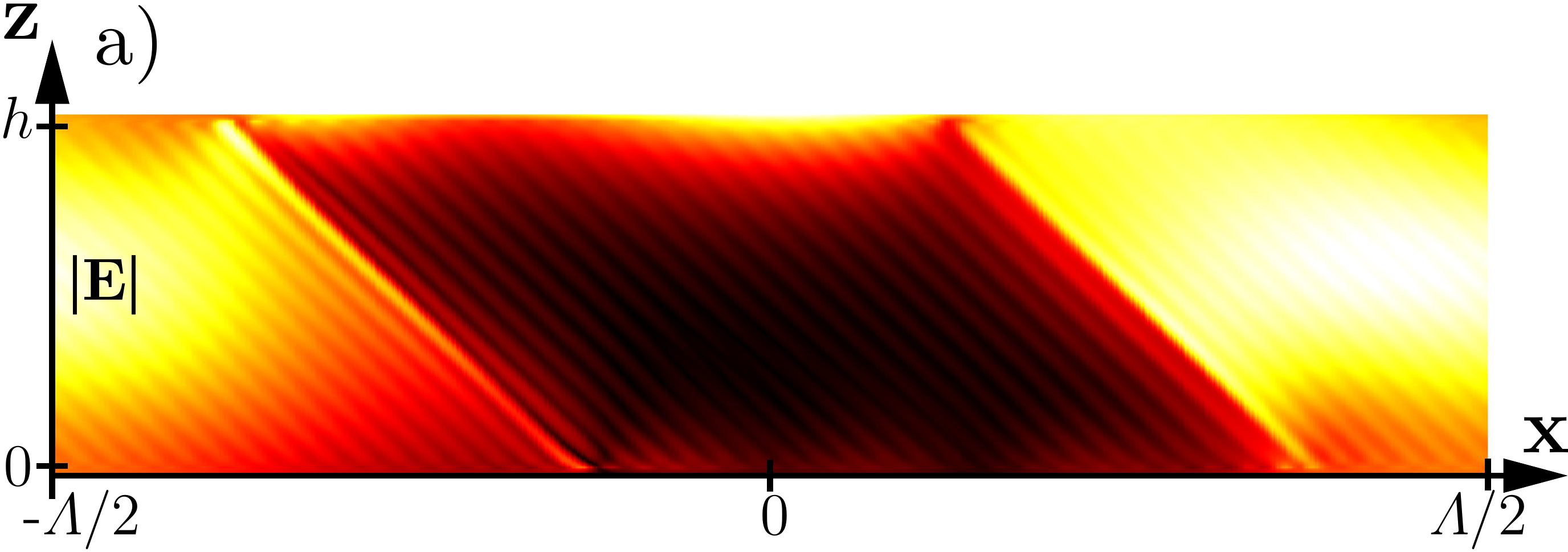}\\
\includegraphics[width=\linewidth]{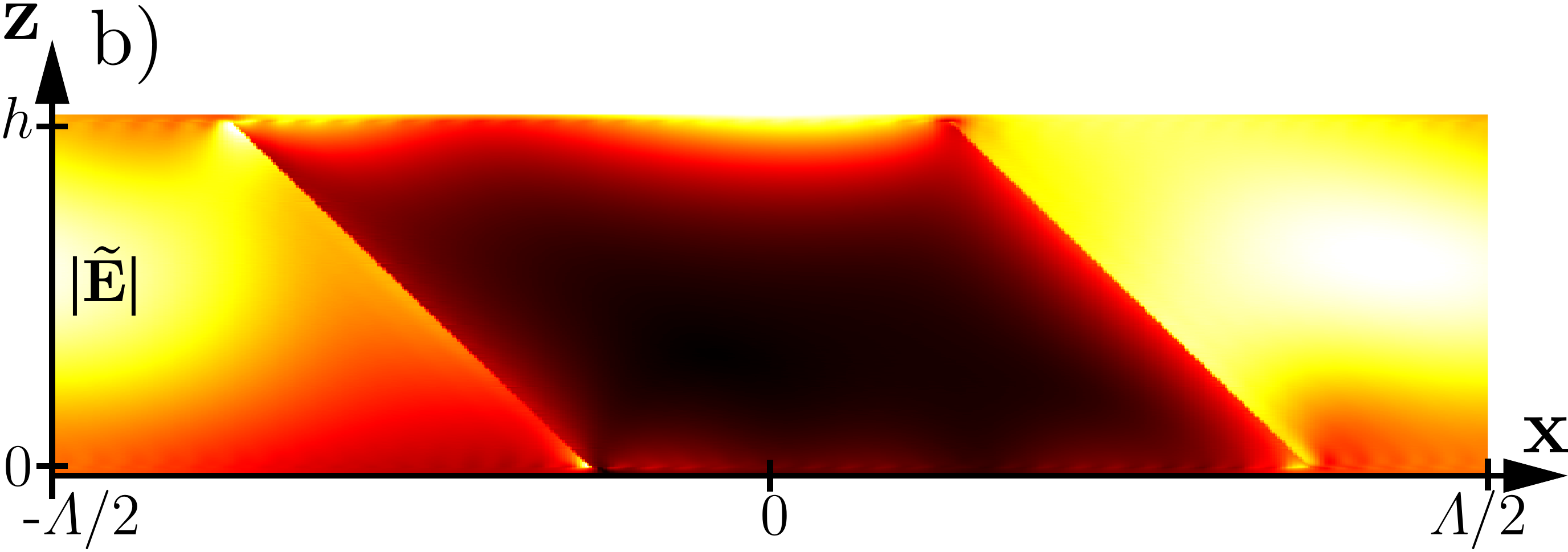}
} \caption{a)
Spatial distribution of the electric near-field, $|E_x^{(40,256)}|$, determined using the
conventional in-plane NVF. b) Spatial distribution of the electric near-field, $|\tilde
E_x^{(40,256)}|$, determined for the same grating parameters as in a) but using the modified
out-of-plane NVF formulation. 
} \label{fig:FieldProfiles15D}
\end{figure}

Numerical results for increasing number of harmonics, $N=2,\ldots,320$, and number of
computational layers, $M=2,\ldots,256$, are presented in \figref{fig:1DzOverview}. It can be
inferred from this figure that the in-plane formulation requires both a high number of FS
coefficients, $N$, and layers, $M$, to achieve convergence to a result of $R^{ref}=0.28788$,
whereas the out-of-plane formulation yields fast convergence to $\tilde
R^{ref}=R^{(320,256)}=0.28837$ with respect to $N$, as per figures~\ref{fig:1DzOverview}(a) and
\ref{fig:1DzOverview}(b), respectively. This behavior is in agreement with the findings
reported in \cite{gt10josaa}. The convergence of the calculated near-field, illustrated in
figures~\ref{fig:1DzOverview}(c) and \ref{fig:1DzOverview}(d), exhibits similar features.
Specifically, the in-plane NVF formulation requires both high $N$ and $M$ to achieve a small
self-error of $\Errsym{\E^{(226,256)}}=4.7\cdot 10^{-2}$, whereas this self-error can already be
achieved with $N=10$, $M=256$ in the out-of-plane formulation. This clearly demonstrates a
drastically improved efficiency to the calculation of the near-field of oblique diffraction
gratings of the approach based on the combination of out-of-plane NVF and the accurate near-field
formulation. The highly improved near-field profile is illustrated in
\figref{fig:FieldProfiles15D}(b), which exhibits no unphysical oscillations near the gold-vacuum
interface. This is in sharp contrast to the conventional field evaluation of the in-plane
formulation (cf. \figref{fig:1DzOverview}(a)), which clearly suffers from spurious oscillations.

\subsection{Analysis of slanted 2D-periodic cylindrical diffraction gratings}
\label{sec:sub2Dz} In this section we investigate the efficiency of using the accurate field
evaluation and the out-of-plane NVF formulation to model a challenging, slanted 2D-periodic
diffraction grating. The grating with periods, $\Lambda_1=\Lambda_2=\Lambda =
\SI{1}{\micro\meter}$, is schematically depicted in \figref{fig:Structures}(d) and consists of a
cylindrical rod with radius $r=\SI{0.3}{\micro \meter}$ and height $h=\SI{0.125}{\micro \meter}$, which is slanted by
$\theta=\pi/4$ along the $x$-axis. Again, only the gold grating is considered in this section.
Finally, the incident plane wave is impinging onto the grating along the normal direction, is
polarized along the $x$-axis, and has a wavelength of $\lambda=\SI{2}{\micro \meter}$.

\begin{figure}[t]
\centering
\minibox{\includegraphics[width=\linewidth]{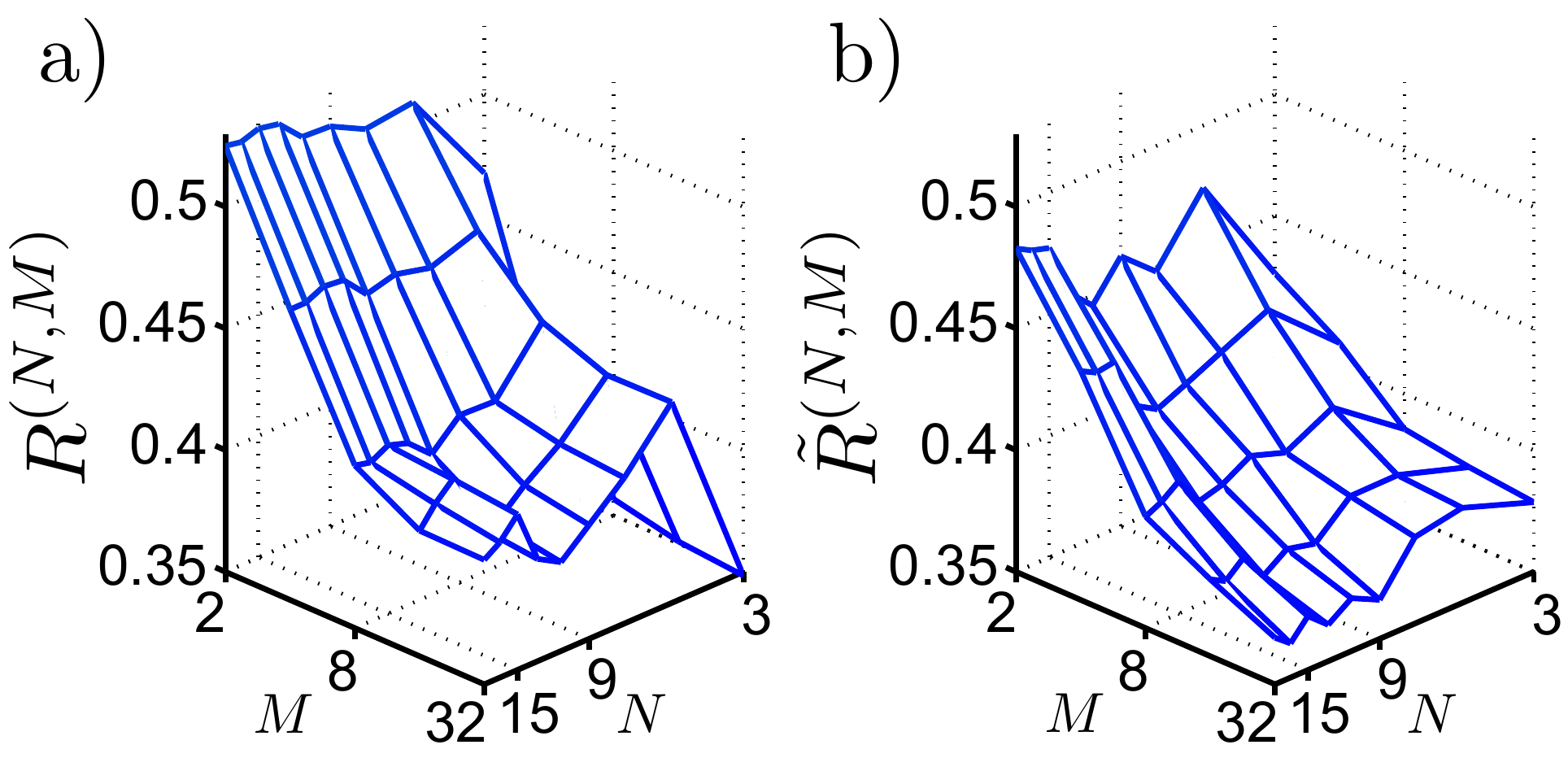} \\
\includegraphics[width=\linewidth]{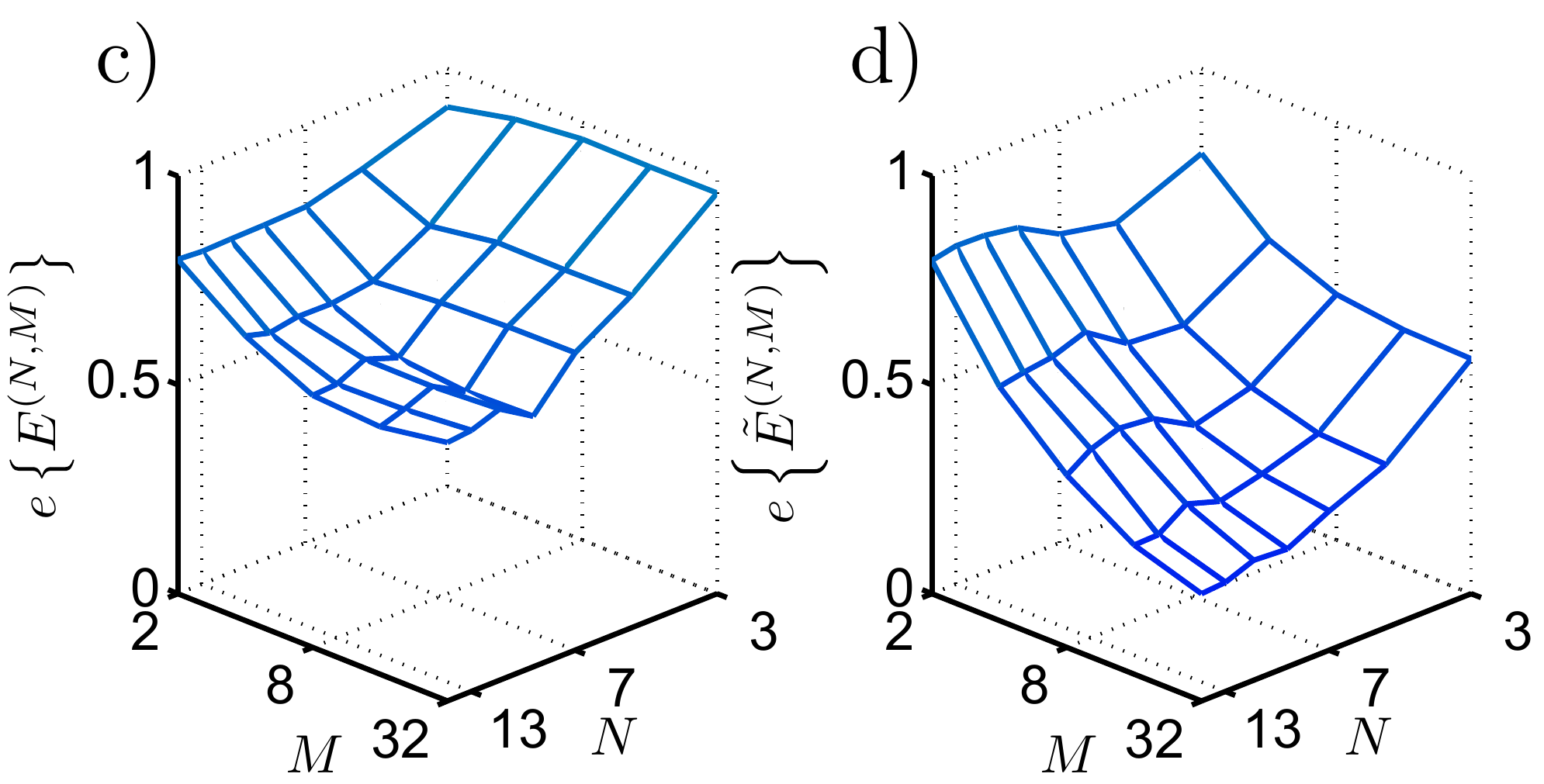} } \caption{Computational results for the slanted 1D binary grating shown in
\figref{fig:Structures}(d). a) Reflection coefficient $R$ vs. $N$ and $M$, determined using the
in-plane NVF formulation. b) Reflection coefficient $\tilde R$ vs. $N$ and $M$, determined using
the out-of-plane NVF formulation. c), d) Near-field convergence of conventional and modified
formulation, respectively.} \label{fig:2DzOverview}
\end{figure}

The reflection coefficient calculated for $N=3,\ldots,19$ harmonics and $M=2,\ldots,32$ layers is
shown in \figref{fig:2DzOverview}(a) and \figref{fig:2DzOverview}(b) and was determined by
using the conventional in-plane NVF and the out-of-plane NVF formulation, respectively. It can be
seen that both approaches converge rather slow, neither one achieve convergence even for the
highest considered values of $M=32$ and $N=19$. Moreover, in order to characterize the error of
the near-field calculations in the two formulations, the self-error with respect to the reference
solutions obtained with $N=19$ and $M=32$ is presented in figures~\ref{fig:2DzOverview}(c) and
\ref{fig:2DzOverview}(d). The in-plane formulation achieves a relative self-error of
$\Errsym{\E^{(13,32)}} = 0.625$, whereas for the same values of $N$ and $M$, the modified field
evaluation in conjunction with the out-of-plane NVF achieves a substantially lower self-error of
$\Errsym{\E^{(13,32)}} = 0.261$.
It has to be stressed, that the necessary accuracy for full convergence could not be achieved in our simulations. The evolution of the computational results for the shown values of $N\leq 19$ and $M\leq 32$ however can be interpreted in favor of the out-of-plane NVF formulation, due to the lower near-field self-error $\Errsym{\E^{(13,32)}} < \Errsym{\E^{(13,32)}}$. It can be supposed that future simulations with finer discretization will reveal the practical benefit of the out-of-plane NVF formulation in conjunction with the modified field formulation for 2D-periodic slanted structures, similar to the case of 1D-periodic slanted diffraction gratings.

\section{Conclusions}
\label{sec:Conclusions} To summarize, we have analyzed the numerical near-field calculated using
the RCWA and identified the Gibb's phenomenon as the main reason for their slow convergence and
the spurious oscillations from which they suffer. As a solution to these deficiencies of RCWA, we
proposed an improvement of this methods that can be applied for modeling arbitrary diffraction
gratings and, more generally, periodic optical structures. The modified formulation significantly
improves the accuracy of 1D-RCWA calculations, for both straight and slanted gratings, where it
speeds up convergence and removes the numerical artifacts from the calculated near-fields. The
accuracy of 2D-periodic grating simulations can be enhanced, however, to a lesser extent than in
the 1D case. The reduced performance in 2D can be attributed to the discontinuity and
non-exactness of the numerical NVF, which is at the core of the modified formulation. Therefore,
it might be fruitful to investigate more elaborate NVF-formulations and their suitability for
near-field calculations, such as a complex valued NVF \cite{av10oe}, which unlike the NVF used
here is continuous everywhere in the grating region.

We expect that the proposed modification of the RCWA method will greatly advance its computational
capabilities, especially for 1D periodic optical structures. In particular, this improved method
could prove instrumental to accurate modeling of periodic plasmonic structures, diffraction
gratings, and surface-nonlinear devices, namely to simulation of physical systems whose
functionality rely on the electromagnetic near-field at interfaces.

\section*{Acknowledgments}
The work of M. Weismann was supported through a UCL Impact Award
graduate studentship funded by UCL and Photon Design Ltd. N. C. P. acknowledges support from
European Research Council / ERC Grant Agreement no. ERC-2014-CoG-648328.

\section*{References}

\end{document}